%% file: main.tex
\documentclass[conference]{IEEEtran}
\ifCLASSINFOpdf
\else
\fi

\usepackage{caption}
\usepackage{subcaption}
\usepackage[flushleft]{threeparttable}
\usepackage{url}
\usepackage{xcolor}
\usepackage{xspace}
\usepackage[many]{tcolorbox}
\usepackage[normalem]{ulem}
\useunder{\uline}{\ul}{}
\usepackage{booktabs}
\usepackage{tabularx}
\usepackage{colortbl}%
\usepackage{pifont}
\usepackage{multirow}
\usepackage{hyperref}
\usepackage{enumitem}
\usepackage{adjustbox}
\usepackage{multirow}
\usepackage[normalem]{ulem}
\useunder{\uline}{\ul}{}
\usepackage{inconsolata}

\definecolor{darkred}{HTML}{860000}
\definecolor{darkteal}{HTML}{005959}
\definecolor{darkpurple}{HTML}{590059}
\definecolor{darkgrey}{HTML}{434343}

\hypersetup{
  colorlinks   = true,    % Colours links instead of ugly boxes
  urlcolor     = blue,    % Colour for external hyperlinks
  linkcolor    = blue,    % Colour of internal links
  citecolor    = blue      % Colour of citations
}

\newcommand{\gptfour}{\textsc{GPT-4}\xspace}
\newcommand{\gptfouro}{\textsc{GPT-4o}\xspace}
\newcommand{\gptfouropre}{\textsc{GPT-4o-2024-08-06}\xspace}

\newcommand{\claude}{\textsc{Claude-3.5-Sonnet}\xspace}
\newcommand{\deepseek}{\textsc{DeepSeek-V3}\xspace}
\newcommand{\llama}{\textsc{Llama3-70b}\xspace}
\newcommand{\gpto}{\textsc{GPT-o1}\xspace}

\newcommand{\tool}{\textsc{IRCopilot}}
\newcommand{\toolgptfour}{\textsc{IRCopilot-GPT-4}}
\newcommand{\toolgptfouro}{\textsc{IRCopilot-GPT-4o}}
\newcommand{\toolclaude}{\textsc{IRCopilot-Claude-3.5-Sonnet}}
\newcommand{\toolgptfouropre}{\textsc{IRCopilot-GPT-4o-2024-08-06}}
\newcommand{\tooldeepseek}{\textsc{IRCopilot-DeepSeek-V3}}
\newcommand{\toolllama}{\textsc{IRCopilot-Llama3-70b}}
\newcommand{\toolgptoone}{\textsc{IRCopilot-GPT-o1}}

\newcommand{\toolnoplanner}{\textsc{IRCopilot-no-Planner}}
\newcommand{\toolnogenerator}{\textsc{IRCopilot-no-Generator}}
\newcommand{\toolnoreflector}{\textsc{IRCopilot-no-Reflector}}
\newcommand{\toolnoanalyst}{\textsc{IRCopilot-no-Analyst}}

\newcommand{\ie}{\textit{i}.\textit{e}.}

\newcommand{\Tref}[1]{Table~\ref{#1}}

\newcommand{\Fref}[1]{Figure~\ref{#1}}
\newcommand{\Sref}[1]{Sec.~\ref{#1}}

\definecolor{lightblue}{RGB}{246,246,246}
\definecolor{darkblue}{RGB}{164,164,164}
\definecolor{lightgreen}{RGB}{251,251,251}
\definecolor{darkgreen}{RGB}{210,210,210}
\definecolor{lightyellow}{RGB}{254,248,245}
\definecolor{darkyellow}{RGB}{247,182,153}
\definecolor{mediumblue}{RGB}{0,102,205}

\newtcolorbox{custombox}[2][]{text width=0.95\linewidth, fontupper=\normalsize,
fonttitle=\bfseries\sffamily\normalsize, enhanced,
attach boxed title to top left={yshift=-2mm,xshift=3mm}, sharp corners,
boxed title style={colframe=black}, top=4pt, bottom=2pt, left=2pt, right=2pt, title=#2, colback=white} 

\newtcolorbox{mybox}[1][]{
    colback=lightblue,
    colframe=darkblue,
    boxrule=0.5mm,
    arc=1mm,
    title=#1,
    fonttitle=\bfseries,
    fontupper=\ttfamily\small, % \ttfamily 现在会指向 Inconsolata 字体
    coltitle=black,
    #1
}

\newtcolorbox{examplebox}[1][]{
    colback=lightgreen,
    colframe=darkgreen,
    boxrule=0.5mm,
    arc=1mm,
    title=#1,
    fonttitle=\bfseries,
    fontupper=\ttfamily\small, % \ttfamily 现在会指向 Inconsolata 字体
    coltitle=black,
    #1
}

\newtcolorbox{codebox}[1][]{
    colback=lightyellow,
    colframe=darkyellow,
    boxrule=0.5mm,
    arc=1mm,
    title=#1,
    fonttitle=\bfseries,
    fontupper=\ttfamily\small, % \ttfamily 现在会指向 Inconsolata 字体
    coltitle=black,
    #1
}

\begin{document}
\title{\tool{}: Automated Incident Response with Large Language Models}


\newcommand{\mkfafu}[0]{{{$^1$}}}
\newcommand{\mkastar}[0]{{{$^2$}}}
\newcommand{\mkntu}[0]{{{$^3$}}}
\newcommand{\mkSanming}[0]{{{$^4$}}}
\newcommand{\mkletter}[0]{{{\normalsize \textsuperscript{\dag}}}}

\author{
    {\rm Xihuan Lin}\mkfafu \rm ,
    {\rm Jie Zhang}\mkastar \rm ,
    {\rm Gelei Deng}\mkntu \rm ,
    {\rm Tianzhe Liu}\mkfafu \rm ,
    {\rm Tianwei Zhang}\mkntu \rm ,
    {\rm Riqing Chen}\mkSanming\mkletter \rm, 
    {\rm and Qing Guo}\mkastar\mkletter \rm\\
    \mkfafu {Independent researchers},\\
    \mkastar {IHPC and CFAR, A*STAR, Singapore},\\
    \mkntu {Nanyang Technological University}\\
    \mkSanming {Sanming University},\\
    \mkletter {\textit{Corresponding Author}}\\

    \medskip
    \textit{h3ppo521@outlook.com}, 
    \textit{\{zhang\_jie, guo\_qing\}@cfar.a-star.edu.sg},\\
    \textit{\{gelei.deng, tianwei.zhang\}@ntu.edu.sg}

}
    

%Telephone: (800) 555--1212, Fax: (888) 555--1212}


\IEEEoverridecommandlockouts
\makeatletter\def\@IEEEpubidpullup{6.5\baselineskip}\makeatother
\IEEEpubid{\parbox{\columnwidth}{
		Network and Distributed System Security (NDSS) Symposium 2025\\
		24-28 February 2025, San Diego, CA, USA\\
		ISBN 979-8-9894372-8-3\\
		https://dx.doi.org/10.14722/ndss.2025.[23$|$24]xxxx\\
		www.ndss-symposium.org
}
\hspace{\columnsep}\makebox[\columnwidth]{}}

\maketitle

% \begingroup\renewcommand\thefootnote{\textsuperscript{\dag}}
% \footnotetext{Corresponding Author}
% \endgroup

\begin{abstract}
Incident response plays a pivotal role in mitigating the impact of cyber attacks. In recent years, the intensity and complexity of global cyber threats have grown significantly, making it increasingly challenging for traditional threat detection and incident response methods to operate effectively in complex network environments. While Large Language Models (LLMs) have shown great potential in early threat detection, their capabilities remain limited when it comes to automated incident response after an intrusion.
To address this gap, we construct an incremental benchmark based on real-world incident response tasks to thoroughly evaluate the performance of LLMs in this domain. Our analysis reveals several key challenges that hinder the practical application of contemporary LLMs, including context loss, hallucinations, privacy protection concerns, and their limited ability to provide accurate, context-specific recommendations.

In response to these challenges, we propose \tool{}, a novel framework for automated incident response powered by LLMs. \tool{} mimics the three dynamic phases of a real-world incident response team using four collaborative LLM-based session components. These components are designed with clear divisions of responsibility, reducing issues such as hallucinations and context loss. Our method leverages diverse prompt designs and strategic responsibility segmentation, significantly improving the system's practicality and efficiency.
Experimental results demonstrate that \tool{} outperforms baseline LLMs across key benchmarks, achieving sub-task completion rates of 150\%, 147\%, 138\%, 136\%, 119\%, and 114\% for various response tasks. Moreover, \tool{} exhibits robust performance on public incident response platforms and in real-world attack scenarios, showcasing its strong applicability.
\end{abstract}


% \fi
\IEEEpeerreviewmaketitle

\input{Tex/1-Introduction}
\input{Tex/2-Background}
\input{Tex/3-ThreatModel}
\input{Tex/4-BenchmarkDesign}

\input{Tex/5-Methodology}
\input{Tex/6-Evaluation}
\input{Tex/7-Discussion}
\input{Tex/8-Conclusion}
\ifCLASSOPTIONcaptionsoff
  \newpage
\fi

\clearpage
\bibliographystyle{IEEEtranS}
\bibliography{main.bib}

\input{Tex/Appendix}

\end{document}

%% file: Tex/1-Introduction.tex
\section{Introduction}

\begin{figure}[t]
  \centering
  \includegraphics[width=0.9\linewidth]{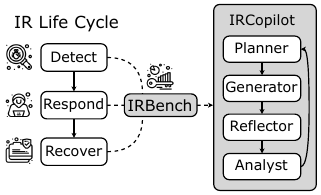}
  % \vspace{-1em}
  \caption{Two main contributions of this work: 1) We build an Incident Response (IR) benchmark to evaluate the contemporary LLMs on IR tasks, in terms of three stages. 2) We propose \tool{} to enhance the performance on IR tasks. }
  \label{fig:intro-overview}
  \vspace{-1em}
\end{figure}

Incident Response (IR) is a cornerstone of system and network security, countering cyber attacks to minimize damage. As illustrated in \Fref{fig:intro-overview}, the IR lifecycle progresses through detection, response, and recovery stages~\cite{nelson2024incident}. Efficient IR requires rapid decision-making, cross-functional teamwork, and adaptability to evolving threats. However, current IR practices, heavily reliant on manual analysis, specialized expertise, and structured collaboration, struggle to address the growing complexity and frequency of cyber threats. Our preliminary survey in~\Sref{sec:background} underscores these challenges, revealing that traditional security operations and response methods are increasingly inadequate. Consequently, enhancing automation in IR has become a critical research topic.

In recent years, the development of Natural Language Processing (NLP)~\cite{chowdhary2020natural} and Large Language Models (LLMs), such as GPT~\cite{brown2020language, achiam2023gpt} and BERT~\cite{devlin2018bert}, offers a promising new frontier for automating security analysis and decision-making. In the field of cybersecurity, LLMs have demonstrated strong potential~\cite{deng2024pentestgpt, wu2024autopt, song2024audit} and are expected to serve as "intelligent assistants" in IR. Some preliminary studies have utilized LLMs to assist in security analysis and threat intelligence extraction~\cite{li2023prompting, alam2023looking, qi2023loggpt}, but they primarily focus on specific tasks in the early stages of threat scenarios, such as intelligence gathering and detection, lacking understanding of and support for comprehensive IR. A critical gap remains: there is no comprehensive approach that leverages the full potential of LLMs across the entire IR lifecycle.

% In highly complex incident response processes with stringent security requirements, LLMs face multiple challenges. These challenges include technical difficulties in contextual understanding and multi-step reasoning, as well as erroneous outputs caused by the phenomenon of "hallucinations". Additionally, practical applications must balance security compliance factors such as privacy protection, the risk of false reporting, and the potential for misuse in attacks. Therefore, further in-depth exploration and research are necessary to address questions such as 1) "How can LLMs help security personnel identify and respond to security incidents more quickly?", 2) "How can the controllability and reliability of LLMs in automated processes be ensured?", 3) "How can cloud-based LLM services be utilized without disclosing sensitive data?", and 4) "How can effective evaluation criteria be established to measure the automated incident response capabilities of LLMs?"

To bridge this gap, a holistic benchmark is urgently needed to evaluate the capabilities of LLMs in incident response. Existing evaluation frameworks remain fragmented and task‑specific. Accordingly, we develop an IR benchmark spanning three mainstream platforms (\textit{i.e.}, XuanJi~\cite{XuanJi}, ZGSF~\cite{ZGSF}, and TryHackMe~\cite{TryHackMe}), comprising a total of 12 case machines and 130 sub-tasks. This benchmark covers the three main stages of the IR lifecycle as defined by NIST 3.0~\cite{nelson2024incident}, with each sub-task mapped to more detailed execution steps within the corresponding guidelines. By systematically assessing the completion rates of sub‑tasks, we can thoroughly characterize the model’s capabilities in IR.

We then conduct an exploratory study using the designed benchmark to evaluate the capabilities of LLMs in IR tasks, with a focus on handling real-world complexities. We test four commercial LLMs (\ie, \gptfour{}~\cite{API}, \gptfouro{}~\cite{API}, \claude{}~\cite{claudeAPI}, and \gpto{}~\cite{API}) and two open-source LLMs (\llama{}~\cite{dubey2024llama} and \deepseek{}~\cite{deepseekAPI}) under controlled conditions with identical environment configurations and prompts. Experiment results show that LLMs cannot effectively overcome these tasks. Specifically, they face the following challenges in IR tasks: (1) difficulty in formulating appropriate response strategies; (2) generation of inaccurate guidance or commands; (3) overlooking critical details; (4) limited long-term memory and poor handling of multiple lines of evidence; (5) potential privacy risks due to exposure to sensitive information.

To enhance automation in incident response, minimize human errors, shorten response time, and provide intelligent decision support during unexpected security incidents, we propose \tool{}, an interactive automated IR system driven by LLMs. 
Targeting the identified challenges, \tool{} consists of four functional components: Planner, Generator, Reflector, and Analyst (see \Fref{fig:intro-overview}), and can be divided into three stages of thought: reasoning, action, and reflection. (1) In the reasoning stage, the Planner is responsible for overall task planning and decomposing complex problems into more detailed sub-tasks. The Analyst evaluates and analyzes the execution results and feeds the analysis conclusions back to the Planner to support subsequent decision-making and optimization. Such decision process effectively overcome the challenges of LLMs to zero-shot the correct response sequence. (2) In the action stage, the Generator produces specific execution plans based on the Planner’s instructions. Since the overall task is decomposed to sub operations individually, the attention/memeory issues of LLMs~\cite{zhang2023language, zhang2023siren} are effectively mitigated. (3) Finally, in the reflection stage, the Reflector integrates multi-source information to review the execution process and propose improvement suggestions, forming a complete closed loop from strategy formulation to continuous optimization. By integrating the aforementioned four LLM components, \tool{} can efficiently support various functions such as incident detection and classification, evidence analysis and threat intelligence correlation, emergency plan formulation, interactive decision-making and execution recommendations, as well as built-in security compliance auditing and data protection. Throughout this process, a seamless connection is achieved from high-level strategy to precise execution and intelligent data analysis, ensuring the coherence and efficiency of IR.
%--------------------------------------------------

We evaluate \tool{} across various testing scenarios to verify its effectiveness and scope of applicability. The results indicate that, in the benchmark tests, \tool{} achieve sub-task completion rates of 1.5$\times$, 1.47$\times$, 1.38$\times$, 1.36$\times$, 1.19$\times$, and 1.14$\times$ higher than directly applying \gptfour{}, \llama{}, \deepseek{}, \gptfouro{}, \claude{}, and \gpto{} respectively, thus demonstrating significantly better performance than the baseline methods. In real-world challenges (\ie, TryHackMe, XuanJi, and ZGSF), \tool{} successfully resolve the majority of IR tasks, underscoring its practical value in enhancing efficiency and accuracy. Furthermore, we conduct an evaluation on five real network attack cases and find that \tool{} effectively identifies and extracts critical traces and information left by attackers. More details are available on Anonymous GitHub~\cite{IRCopilot}.

% \textcolor{blue}{As a long-term research objective, we are committed to fully leveraging advanced machine learning methods and developing an entirely automated incident response system to improve the utilization of LLMs on the CTI (Cyber Threat Intelligence) lifecycle~\cite{chen2024survey, nova2022security}.} The overall architecture, as shown in Figure~\ref{fig:introoverview}, illustrates our current work in the CTI field and contributions to future development plans.

To summarize, we make the following key contributions:

\begin{itemize}[leftmargin=*,noitemsep,topsep=0pt]

% \item \textbf{Comprehensive Survey of Recent Incident Response Trends and Cybersecurity Landscape.} This survey reveals numerous findings they are currently encountering and highlights the limitations of existing solutions.

\item \textbf{Development of a Comprehensive Incident Response Benchmark.} Our benchmark covers numerous target machines from prominent platforms (TryHackMe, XuanJi, and ZGSF). It comprises 12 representative targets (with 130 sub-tasks) and spans key stages of IR, drawing partly on real-world scenarios. As far as we know, this is the first comprehensive benchmark specifically for IR, featuring phased evaluation and comparison capabilities. Furthermore, we evaluate the performance of multiple LLMs on IR within this benchmark and analyze the key challenges they face.

\item \textbf{Design of an Innovative LLM-Based Incident Response System.} Our automated and interactive system, \tool{}, provides efficient and innovative IR capabilities. Through various technological innovations, it significantly boosts IR accuracy and efficiency while minimizing its impact on the system, ensuring overall stability and practicality.

\item \textbf{Systematic Evaluation and Real-World Validation.} The evaluation results indicate that \tool{} based on the \gptfouro{} series successfully achieve objectives in the vast majority of tasks across complex environments, demonstrating its practicality and reliability. Furthermore, validation in real-world scenarios across two major OSs demonstrates the versatility and effective applicability of this approach to IR.

\end{itemize}

%% file: Tex/2-Background.tex
\section{Background \& Related Work}\label{sec:background}
\subsection{Global Cybersecurity Landscape}

In recent years, the global cybersecurity ecosystem has been facing unprecedented challenges of complexity and dynamism. We conduct an in-depth analysis of over 200 annual reports on cybersecurity or IR, and draw the following observations, with detailed data disclosed in the Appendix Figures~\ref{fig:CVEs}, \ref{fig:Weekly-Attacks}, \ref{fig:Attack-Vectors}, \ref{fig:Investigation-Types} and our anonymous GitHub~\cite{IRCopilot}.

\noindent\textbf{(1) Expanding Attack Surface and Rising Vulnerabilities}. The growing complexity and interconnectivity of software systems continuously expand the attack surface, increasing the number and severity of vulnerabilities. Per NVD data~\cite{NVD}, published CVEs surged from \textasciitilde7,928 in 2014 to 40,287 in 2024, showing an exponential rise. The cumulative CVE count grew from under 70,000 to over 262,000 in a decade.

\noindent\textbf{(2) Diversification of Vulnerability Types and Cybersecurity Incidents.} These vulnerabilities are not confined to a single exploitation technique but exhibit highly diversified and specialized characteristics. A major concern is that the proportion of high-risk vulnerabilities has risen sharply.

\noindent\textbf{(3) Prominence of Industry and Scenario Specificity}. Attacks on critical sectors like finance, government, education, and healthcare are increasing in frequency and sophistication, as shown by Check Point~\cite{checkpoint} from 2020 to 2023. Industry-tailored attack patterns are increasingly sophisticated.

\noindent\textbf{(4) Comprehensive Upgrade of Cybersecurity Requirements}. IR studies reveal evolving threats, with Unit 42~\cite{unit42} data (2021--2023) highlighting ransomware, BEC, network intrusions, and web breaches as primary concerns. Attackers' yearly tactical shifts increase IR teams' workload and error risks. Kaspersky~\cite{kaspersky} data links attack vectors to economic losses, data breaches, service disruptions, and ethical issues.

\subsection{Incident Response}\label{sec:background:ir}

Cybersecurity incident response involves organized, systematic actions taken by an organization upon detecting a security incident or potential breach due to unauthorized activities. These activities may compromise the confidentiality, integrity, or availability of systems or assets, or violate laws, policies, or security protocols. The aim is to reduce harm, limit losses, and quickly restore normal operations. 
As shown in \Fref{fig:intro-overview}, the lifecycle of IR consists of three key stages~\cite{nelson2024incident}: Detect, Respond and Recover. In the Detect phase, potential cybersecurity threats and compromises are identified and analyzed. The Respond phase analyzes the system and takes necessary measures to address detected cybersecurity incidents. Finally, in the Recover phase, efforts are made to restore affected assets and operations to their normal state.

\noindent\textbf{Limitations.} The increasing persistence, stealth, and unpredictability of cyber threats challenge traditional threat detection methods. Conventional rule-based and machine learning methodologies excel in a priori tasks, such as real-time intrusion detection and threat analysis, due to their low latency and precision~\cite{yuan2021deep}. However, these approaches stumble when it comes to incident response, particularly in post-incident forensic analysis and adaptive mitigation. These methods rely on fixed rules and pre-trained models, which hampers their ability to keep up with an ever-changing threat landscape, leaving them less suited to the dynamic, reactive demands of IR. In the past, IR was a labor-intensive process, relying heavily on manual efforts and expert knowledge, with teams laboriously investigating the overall scope of incidents, analyzing threats, and manually resolving critical issues within tight time constraints. However, as threats grow more sophisticated, purely manual approaches struggle to deliver timely, efficient outcomes. Traditional a priori methods, which can support initial assessments like network intrusion detection~\cite{fu2023detecting}, focus on specific tasks and lack the scalability needed to address the full spectrum of IR. In addition, effective IR requires consolidating fragmented information from diverse sources into a unified situational picture. Consequently, integrating artificial intelligence to automate IR remains a significant challenge. Moreover, cyber threats do not just exploit system vulnerabilities, they can also stem from diverse sources like internal actors and espionage, making analysis even trickier. When faced with massive amounts of network data and unpredictable scenarios, security researchers require more adaptable tools. LLMs show promise here, leveraging their flexibility and contextual understanding~\cite{chen2024survey}. 

\subsection{Large Language Models and Their Applications for Cybersecurity}
With the development of LLMs recently, significant transformations have occurred across various fields. Renowned models such as GPT~\cite{brown2020language, achiam2023gpt}, Llama~\cite{dubey2024llama}, Claude~\cite{Claude}, and DeepSeek~\cite{deepseek} have demonstrated excellent performance in numerous domains, including text generation~\cite{li2024pre}, translation~\cite{wang2019learning}, programming~\cite{li2023large}, and summarization~\cite{zhang2024benchmarking}. 
% Notably, OpenAI's GPT series, especially \gptfour{} and \gptfouro{}, have become the foundational models for our exploration of automated cybersecurity incident response due to their powerful natural language processing capabilities.

To further enhance the performance of LLMs in complex task scenarios, researchers have introduced and adopted various cutting-edge planning and reasoning methods, collectively known as prompt engineering. These methods include CoT (Chain of Thought)~\cite{wei2022chain}, ToT (Tree of Thoughts)~\cite{yao2024tree}, RAG (Retrieval Augmented Generation)~\cite{lewis2020retrieval}, ReAct (Reasoning and Acting)~\cite{yao2023react}, and Reflexion~\cite{shinn2024reflexion}. They enable LLMs to generate more factually grounded responses and perform better in intricate analytical tasks. For instance, ToT constructs a thought tree structure, prompting the model to engage in multi-level reasoning before providing the final answer, thereby improving the accuracy and reliability of responses. These characteristics make ToT particularly suitable for the deep analysis and decision-making required in automated cybersecurity IR. We will investigate how to employ these prompt engineering techniques to optimize the application effectiveness of LLMs in automated IR.

LLMs, with their deep contextual understanding and reasoning capabilities, have demonstrated considerable potential in numerous areas of cybersecurity, including code analysis~\cite{pearce2022asleep}, penetration testing~\cite{deng2024pentestgpt, wu2024autopt}, OSINT (Open Source Intelligence)~\cite{shafee2024evaluation}, vulnerability remediation and management~\cite{pearce2023examining, liu2024exploring}, malicious webpage identification~\cite{li2023prompting}, attack pattern extraction from CTI (Cyber Threat Intelligence)~\cite{alam2023looking}, log-based anomaly detection~\cite{qi2023loggpt}, and developing privacy-preserving models for IoT~\cite{ferrag2023revolutionizing}. However, existing research focuses mainly on early threat stages, such as intelligence gathering and detection. With the escalating severity of cyber threats, the domain of automated IR post-intrusion remains relatively underdeveloped. To address this gap, we propose an LLM-based automated incident response framework. Furthermore, there is currently a lack of a benchmark capable of evaluating the performance of LLMs in IR. Our research will establish a professional, objective, and representative benchmark to provide a reliable reference for subsequent studies in this field.

% However, with the increasing severity of cyber threats and the enhancement of automated attack trends, the field of automated incident response for intrusion detection remains relatively underdeveloped. To this end, our work aims to propose an automated incident response framework based on LLMs to better address complex cyber situations. Furthermore, there is currently no benchmark to evaluate the capabilities of LLMs in incident response, thus we need to construct a professional, objective, and representative evaluation benchmark.

%% file: Tex/3-ThreatModel.tex
\section{Threat Model}\label{sec:threatmodel}
We consider a practical scenario with three parties: a victim, an attacker, and a third-party IR service provider (responder).

\noindent\textbf{Victim's Capabilities \& Goals.} 
Operating from a white-box perspective, the victim's IT infrastructure is equipped with an intrusion detection system (IDS) capable of identifying unauthorized access or anomalous network activity. Upon detecting a potential threat, the victim promptly informs the responder. Their goal is to minimize attack impact while preserving critical business operations.
% Their goal is to assist in mitigating the attack's impact while ensuring the availability of critical business operations.
% The victim has intrusion detection capabilities and knows the resource is compromised. Upon detecting suspicious activity, they promptly notify the responder, who must quickly intervene, launch an incident response, assess risks, and mitigate the impact of the attack.

\noindent\textbf{Attacker’s Capabilities \& Goals.} 
From a gray-box perspective, the attacker seeks to steal sensitive data, extort profit, or disrupt operations by gaining unauthorized access. Their capabilities include: 
(1) adequate resources (e.g., computing power, libraries) and experience in multi-stage attacks and evasion techniques. 
(2) partial knowledge of the victim's security defenses, allowing bypass of specific protections. 
Furthermore, they (3) inevitably leave traces in the compromised system.

% The attacker achieves specific objectives like stealing sensitive information, extorting for profit, or causing other serious consequences by attacking IT resources and gaining unauthorized access. Typically, they implement penetration and control of target systems or networks according to pre-established plans, leaving objective traces of intrusion within systems or other network devices during the process. We assume that attackers possess the following conditions: (1) adequate critical resources to support long-term, large-scale, or distributed attack operations; (2) the ability to employ diversified and multistage attack strategies and exploit libraries; and (3) familiarity with the security defenses within the target systems or networks, along with certain evasion or counteracting capabilities. However, attackers also: (4) inevitably leave traces in the intruded system.

\noindent\textbf{Responder's Capabilities \& Goals.} 
The responder operates from a gray-box perspective, possessing: 
(1) limited privileges to access victim data, restricted to information necessary for IR. 
(2) partial knowledge of attacker traces and compromised assets. 
Their responsibility to respond effectively while minimizing disruptions to the victim's core business operations.

% \noindent\textbf{Responder’s Capabilities \& Goals.} 
% The Responder is a trusted third party, offering cybersecurity incident response services to its clients. Based on its role and authorized scope, the service provider (1) possesses limited privileges for data from the victim, ensuring that no sensitive information unrelated is involved; and (2) retains partial knowledge of attackers' residual traces within the system or network, as well as the critical points that have been compromised. However, the service provider (3) must maintain the availability of core business operations and key resources, avoiding unnecessary disruptions or damage to normal operations.

%% file: Tex/4-BenchmarkDesign.tex
\section{IRBench: A Benchmark for Incident Response}\label{sec:benchmark}
\subsection{Motivation}\label{sec:benchmark:motivation}
% The evaluation of LLMs in the context of IR necessitates an objective and comprehensive benchmark. Existing benchmarks in this domain~\cite{IncidentResponseMetrics, 555Benchmark, force2017security}, while valuable, exhibit significant limitations when applied to LLM assessment. These benchmarks primarily emphasize the completion of overarching task workflows, often neglecting the detailed evaluation of specific sub-tasks and intermediate steps within the IR process. Although such macro-oriented benchmarks help determine the success or failure of the final response, they fail to adequately reflect the progressive advancements and value accumulation at each stage of the IR. Consequently, these benchmarks struggle to accurately measure the comprehensive capabilities of LLMs in the details and steps of IR, and they cannot provide strong support for precise performance evaluation and optimization improvements of the models.

The absence of suitable benchmarks for evaluating LLMs in IR leaves their reliability and practicality as significant unresolved research questions. We notice several benchmarks exist within the broader cybersecurity domain, for instance, CTIBench~\cite{NEURIPS2024_5acd3c62} established five metrics to measure LLM proficiency in understanding threat intelligence and vulnerabilities. SEvenLLM-Bench~\cite{ji2024sevenllm} employs multiple-choice questions to focus on CTI comprehension, specifically on extracting structured information from security events and analyzing reports to identify attack means and methods for summarization. PentestGPT-Excalibur~\cite{deng2024pentestgpt} introduced the first benchmark in penetration testing, designed for progressive accomplishment assessments and comparisons. However, these benchmarks remain domain-specific and do not address the detailed challenges of real-world IR processes, as they lack appreciation for their inherent complexity and interdependencies. We aim to develop a comprehensive benchmark that assesses LLM performance across the progressive micro-stages of IR, capturing the cumulative value at each step. This benchmark will enable precise evaluation and optimization of LLMs in IR and serve as a valuable resource for cybersecurity professionals. Therefore, the benchmark should include comprehensive and realistic tasks, incorporate a difficulty gradient in its design, and implement phase-oriented effectiveness evaluation.

\begin{table*}[!t]
	\centering
	\caption{Performance of contemporary LLMs on IRBench.}
	\vspace{-0.5em}
	% \vspace{-1em}
	\fontsize{14}{16}\selectfont
	\resizebox{\linewidth}{!}{
		\begin{tabular}{c|cccc|cccc} 
			\toprule
			\multicolumn{1}{c}{}   & \multicolumn{4}{c}{\textbf{Overall}} & \multicolumn{4}{c}{\textbf{Sub-task}} \\ \midrule
			LLMs                   & Easy (3)          & Medium (6)          & Hard (3)          & Total (12)            & Easy (36)           & Medium (62)           & Hard (32)          & Total (130)             \\ \midrule
			\gptfour{}             & 0 (0.00\%)        & 1 (16.67\%)         & 0 (0.00\%)        & 1 (8.33\%)            & 22 (61.11\%)        & 39 (62.90\%)          & 15 (46.88\%)       &  \textbf{76 (58.46\%)}           \\
			\llama{}               & 0 (0.00\%)        & 1 (16.67\%)         & 0 (0.00\%)        & 1 (8.33\%)            & 22 (61.11\%)        & 43 (69.35\%)          & 18 (56.25\%)       &  \textbf{83 (63.85\%)}           \\
			\deepseek{}            & 0 (0.00\%)        & 1 (16.67\%)         & 0 (0.00\%)        & 1 (8.33\%)            & 20 (55.56\%)        & 43 (69.35\%)          & 22 (68.75\%)       &  \textbf{85 (65.38\%)}           \\
			\gptfouro{}            & 1 (33.33\%)       & 1 (16.67\%)         & 1 (33.33\%)       & 3 (25.00\%)           & 28 (77.78\%)        & 45 (72.58\%)          & 21 (65.63\%)       &  \textbf{94 (72.31\%)}           \\
			\claude{}              & 2 (66.67\%)       & 2 (33.33\%)         & 1 (33.33\%)       & 5 (41.67\%)           & 31 (86.11\%)        & 50 (80.65\%)          & 27 (84.38\%)       & \textbf{108 (83.08\%)}          \\
			\gpto{}                & 1 (33.33\%)       & 3 (50.00\%)         & 1 (33.33\%)       & 5 (41.67\%)           & 31 (86.11\%)        & 54 (87.10\%)          & 28 (87.50\%)       & \textbf{113 (86.92\%)}          \\ \midrule
			\textbf{Average}       & \textbf{0.7 (22.33\%)} & \textbf{1.5 (25.00\%)} & \textbf{0.5 (16.66\%)} & \textbf{2.7 (22.25\%)} & \textbf{25.7 (71.31\%)} & \textbf{45.7 (73.66\%)} & \textbf{21.83 (68.2\%)} & \textbf{93.2 (71.67\%)} \\ \bottomrule
		\end{tabular}
	}
	\label{tab:llm_performance}
	% \vspace{-1em}
\end{table*}

\begin{table}[!t]
  \centering
  \caption{Failure Reasons of LLMs on IRBench.}
  \vspace{-0.5em}
  % 在大括号里先切换字体尺寸，再做宽度缩放
  {%
    \fontsize{14}{16}\selectfont  % 字体 12pt，行距 14pt，可根据需要调整
    \resizebox{\linewidth}{!}{%
      \begin{tabular}{l|*{6}{c}|c}
        \toprule
        \textbf{Failure Reasons}       & \textbf{GPT4} & \textbf{Llama} & \textbf{DeepSeek}  & \textbf{4o}  & \textbf{Claude}   & \textbf{o1}  & \textbf{Total}\\
        \midrule
        False IR Strategy              & 20            & 23             & 20                 & 14           & 12               & 8             & \textbf{97}   \\
        False Command Generation       & 11            & 8              & 6                  & 10           & 2                & 1             & \textbf{38}   \\
        Key Information Ignored        & 7             & 4              & 7                  & 5            & 1                & 4             & \textbf{28}   \\
        False Guidance Generation      & 8             & 4              & 4                  & 2            & 2                & 4             & \textbf{24}   \\
        False Result Interpretation    & 5             & 7              & 5                  & 2            & 2                & 0             & \textbf{21}   \\ 
        Session context lost           & 3             & 1              & 3                  & 3            & 3                & 0             & \textbf{13}   \\
        \midrule
        Total                          & 54            & 47             & 45                 & 36           & 22               & 17            & \textbf{221}  \\
        \bottomrule
      \end{tabular}%
    }%
  }%
  \label{tab:llm-failure}
  % \vspace{-2em}
\end{table}

\subsection{Benchmark Design}\label{sec:benchmark:design}
% 修改完成

Guided by the criteria above, we introduce IRBench, a benchmark designed to systematically map real‑world IR tasks. A panel of three anonymous experts contributes to the design. It comprises an independent practitioner and two members of the research team, each with at least five years of experience in incident response and vulnerability analysis, CISSP certification, and a record of CVE discoveries.

% 修改完成
% \noindent\textbf{Task Selection.} In the task selection process, we focus on simulated environments grounded in real-world cybersecurity incidents. We curate tasks from mainstream cybersecurity practice platforms (\ie, XuanJi~\cite{XuanJi}, ZGSF~\cite{ZGSF} and TryHackMe~\cite{TryHackMe}). Specifically, we rigorously reviewed and manually tested all relevant tasks and virtual environments across each platform to ensure comprehensive coverage and high quality, thereby encompassing a broad spectrum of real-world incident types. From this pool, we selected a representative subset of tasks spanning various incident scenarios and response methods. 
\noindent\textbf{Task Selection.} 
We prioritize simulated environments derived from authentic cybersecurity incidents. IR tasks are curated from XuanJi~\cite{XuanJi}, ZGSF~\cite{ZGSF}, and TryHackMe~\cite{TryHackMe}. To ensure breadth and quality, all IR‑related tasks and virtual machines were rigorously reviewed and manually tested. From this pool, we selected a representative subset of tasks that spans diverse incident scenarios and response techniques. 
% 修改完成
% To simulate real‑world IR scenarios of varying complexity, experts independently performed manual testing of each task, reviewed platform‑wide completion rates, and evaluated them against difficulty standards (including attack vectors and scope of impact).  After group discussion, the tasks were assigned to one of three levels: Easy, Medium, and Hard. Specifically, we perform manual evaluations based on the difficulty levels and completion rates provided by the platform. Each expert assigns scores independently; when the difference between the highest and lowest scores does not exceed four points, we compute the arithmetic mean. If discrepancies are larger, decisions are made by majority vote, with further discussion as necessary to reach consensus.
To reflect real‑world IR complexity, three experts independently tested each task, reviewed platform‑wide completion rates, and scored them against difficulty criteria (e.g., attack vectors, impact scope) provided in Appendix~\Tref{tab:difficulty_levels}. When their ratings differed by no more than four points, we used the arithmetic mean; larger discrepancies were settled by majority vote, with further discussion as needed. As a result, tasks were organized into three levels—Easy, Medium, and Hard—to indicate increasing complexity. 
% 修改完成
Crucially, we prioritize tasks unlikely to appear in LLM training corpora by selecting only those published after standard training cutoffs and excluding widely publicized or duplicated content. This procedure mitigates data contamination and ensures an equitable evaluation of LLM generalization. Further details are discussed in~\Sref{sec:discussion}. All selected tasks exclude real-world sensitive or private data, thereby avoiding ethical and legal concerns.

\noindent\textbf{Task Decomposition.} 
To ensure the rigor and systematic nature of our research methodology, we break down each IR task into a series of sub-tasks across three steps: clue acquisition, information analysis, and incident handling and recovery. The entire process strictly adheres to the guiding principles of NIST SP 800-61 Rev. 3~\cite{nelson2024incident}, mapping each sub-task to the specific step breakdowns defined in the guidelines. Ultimately, we develop a comprehensive list of sub-tasks for each benchmark task, which is provided in Appendix~\Tref{appendix:benchmark}.

\noindent\textbf{Benchmark Validation.} 
To ensure the scientific validity and effectiveness of the benchmark, we implement a rigorous evaluation and iterative optimization process. Specifically, the experts independently complete the assigned benchmark tasks and document a detailed IR process based on their performance and results. Based on the experts' feedback, we selectively adjust the task decomposition strategy, focusing on addressing potential multiple response paths within the tasks, thereby enhancing the applicability and flexibility of the benchmarks in practical applications.

Through the process outlined above, we develop a comprehensive benchmark. IRBench includes 12 IR objectives, each assigned varying difficulty levels, and is further subdivided into 130 specific sub-tasks across 27 categories. These objectives and sub-tasks fully represent the essential technologies, skills, and scenarios integral to IR. Designed with careful attention to task quantity and scope, the benchmark accommodates diverse real-world applications while effectively evaluating multi-level IR capabilities. Detailed content, sub-task classifications, and task-difficulty criteria are available on our anonymous GitHub~\cite{IRCopilot} for reference. Furthermore, publicly available high-difficulty cases are currently limited, and we will continuously expand IRBench to enhance its adaptability.

%---------------------------------------------------------------------------------
\subsection{Evaluation on Contemporary LLMs}\label{sec:benchmark:evaluation}

% We launch a preliminary exploration to evaluate LLMs' capabilities in IR tasks on IRBench, using the human-in-the-loop strategy~\cite{deng2024pentestgpt, hilp, takerngsaksiri2024human, amirizaniani2024developing}. The human expert acts solely as an executor, adhering to the LLM’s instructions without offering independent insights or decisions. We assess the performance of three state-of-the-art commercial LLMs (\gptfour{}, \gptfouro{}, and \claude{}), along with two open-source LLMs (\llama{} and \deepseek{}) and the reasoning-focused LLM (\gpto{}). To ensure fair comparisons, all experiments are performed under consistent conditions, employing identical prompts and environmental settings. The LLMs are explicitly instructed not to use external automated tools as the purpose is to evaluate LLMs' inherent knowledge and capabilities. An incident response consists of multiple interconnected tasks spanning the detection, response, and recovery phases. \textit{Success is only achieve when all sub-tasks in these phases are completed successfully}.

We conduct a preliminary exploration on IRBench to evaluate LLMs’ capabilities in IR tasks via the human-in-the-loop strategy~\cite{deng2024pentestgpt, hilp, takerngsaksiri2024human, amirizaniani2024developing}. In this setup, the human expert serves strictly as an executor of the model’s instructions, without injecting independent judgments. We evaluate three leading commercial LLMs (\gptfour{}, \gptfouro{}, and \claude{}), two open-source models (\llama{} and \deepseek{}) and one reasoning-focused LLM (\gpto{}). All experiments were conducted under standardized conditions, using identical prompts and controlled environments to guarantee an equitable comparison. To isolate each model’s intrinsic reasoning and knowledge, we explicitly prohibited the use of any external tools. An incident response consists of multiple interconnected sub-tasks spanning the detection, response, and recovery phases. \textit{Success is achieved only when all sub-tasks across these phases are completed successfully.}

\begin{figure*}[!ht]
  \centering
% \vspace{-1em}
  \includegraphics[width=\linewidth]{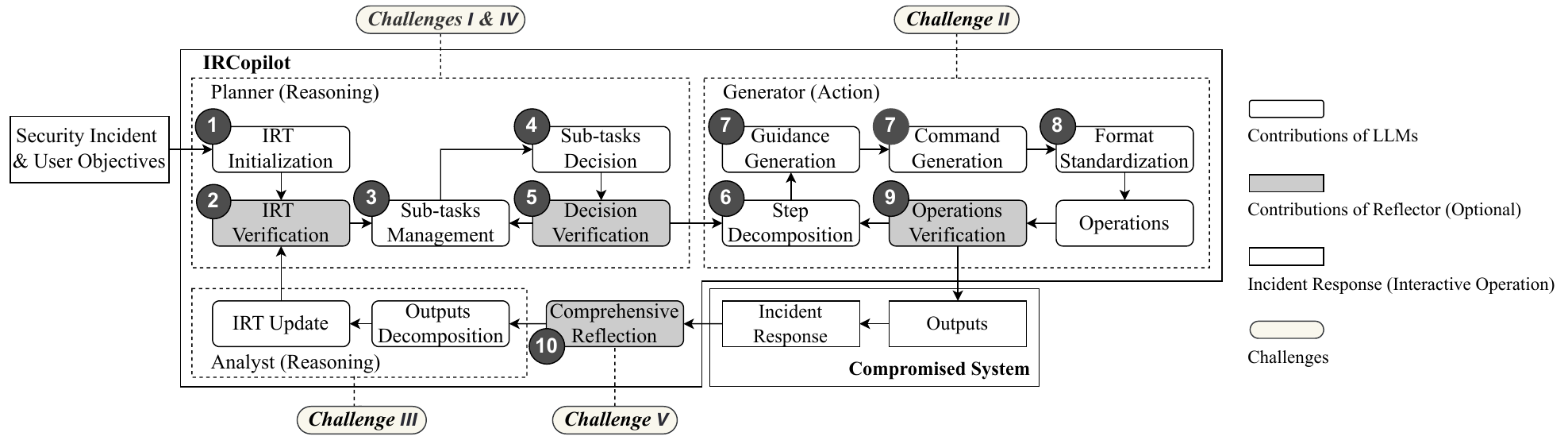}
  % \vspace{-2em}
  \caption{Workflow of \tool{}. This figure illustrates the principle of \tool{}, structured into three cognitive stages: Reasoning, Action, and Reflection. \textbf{1.} In the Reasoning phase, we design the Planner and Analyst to maintain the IRT and tackle \textit{Challenges I, III, and IV} mentioned in~\Sref{sec:benchmark:evaluation} through step-by-step reasoning. \textbf{2.} In the Action phase, the Generator distributes and generates guidance and commands to mitigate \textit{Challenge II}. \textbf{3.} Finally, in the Reflection phase, the Reflector addresses \textit{challenges posed by hallucination and privacy risks}.}
  \label{fig:workflow}
  % \vspace{-1em}
\end{figure*}

\Tref{tab:llm_performance} illustrates the performance of the evaluated LLMs on IRBench by overall task difficulty and by individual sub-tasks. Both \gpto{} and \claude{} solve 5 of the 12 overall tasks, but \gpto{} attains the highest sub-task completion rate at 86.92\%, compared with 83.08\% for \claude{}. In contrast, \gptfouro{}, \deepseek{}, \llama{}, and \gptfour{} achieve lower sub-task rates of 72.31\%, 65.38\%, 63.85\%, and 58.46\%, respectively. All LLMs consistently excel in easy and medium-difficulty tasks, reflecting their strengths in basic analysis and information processing. However, their performance exhibits clear declines on harder tasks, revealing limitations in handling complex IR scenarios. We find that the success of LLMs in IR hinges not only on the overall difficulty of the task but also on the accurate completion of each constituent sub-task, where even slight missteps can cascade into failure. In complex scenarios, performance bottlenecks often stem from weaknesses in specific sub-tasks, which ultimately dictate the entire IR outcome. A single misinterpretation of one requirement can propagate through subsequent steps and jeopardize the result.

% We find that the success of LLMs in IR is not only closely related to the overall difficulty of the task, but also tightly connected to the completion of individual sub-tasks within it, where even subtle differences can precipitate failure. Particularly in complex scenarios, the capabilities of LLMs are frequently constrained by their performance in specific sub-tasks, which ultimately determines the overall IR outcome. For instance, they may falter if it misinterprets a single sub-task requirement, amplifying the impact of such errors in contexts.

% To better understand the limitations, we manually investigate and summarize the causes behind the models’ sub-task failures, with results detailed in \Tref{tab:llm-failure}. 
To gain deeper insight into the limitations of LLMs in IR, we manually review each experiment in a detailed and systematic manner, and systematically classify the causes behind the models’ sub-task failures, with results detailed in \Tref{tab:llm-failure}. The most prevalent failure modes are the inability to formulate effective IR strategies and inaccurate or incomplete guidance. Representative cases illustrating these two dominant issues are given in Appendix~\ref{appendix:case1} and Appendix~\ref{appendix:case2}. Through manual analysis, we pinpoint the specific reasons driving these failures and outline the key challenges below:

% Common failure modes include the inability to devise effective IR strategies and to deliver accurate guidance, etc. To elucidate the most critical issues, we provide concrete examples of the two primary causes of failure in Appendix~\ref{appendix:case1} and~\ref{appendix:case2}. Through manual analysis, we pinpoint the specific reasons driving these failures and outline the key challenges below:

\begin{enumerate}[leftmargin=*,noitemsep,nolistsep,topsep=0pt]
\item \textbf{Difficulty in Strategy Formulation}: LLMs often struggle to develop effective IR strategies. For example, when lacking information they may fail to set a clear direction, and when faced with a large volume of data they cannot synthesize it into optimal next‐step decisions. This limitation stems from challenges in organizing and reasoning over diverse inputs into coherent, actionable plans.

% \item \textbf{Inaccurate Guidance or Commands}: Models occasionally make factually incorrect or suboptimal decisions, and generate commands containing subtle errors or incompatible with the target operating system, ultimately causing failures in IR.
\item \textbf{Guidance and Commands Accuracy Limitations}: Even when a strategy is formulated, LLMs can generate factually incorrect guidance or syntactically flawed commands that are incompatible with the target system. Such subtle errors often cascade into execution failures.

\item \textbf{Overlooking Critical Details}: Over extended, multi‐step workflows, LLMs tend to overlook minor but pivotal clues—such as anomalous log entries or configuration nuances—that are essential for successful resolution.
% During prolonged tasks, LLMs often miss small yet critical details essential for task completion.

% \item \textbf{Limited Long-Term Memory}: LLMs face challenges in maintaining continuity across multi-step IR, integrating and tracking multiple streams of evidence, and preserving coherent context over extended operations.
\item \textbf{Context Retention Constraints}: Maintaining continuity across multi‐stage response processes poses a challenge. LLMs struggle to track parallel evidence streams, integrate new findings with prior context, and preserve a coherent “state” over lengthy interactions.

% \item \textbf{Privacy and Security Risks}: Even when an IR completes successfully, LLMs may unintentionally process or disclose sensitive information during task execution, thereby introducing substantial risks to both privacy and security.
\item \textbf{Privacy and Security Risks}: During analysis, LLMs may inadvertently ingest or expose sensitive data—potentially leaking confidential information or violating compliance requirements—which introduces significant privacy and operational security concerns.
\end{enumerate}

% \textcolor{blue}{Considering (1) ranking them in the sequence of number of failures (or impact the task success). (2) for the most important one or two reasons, add sentences like "we provide a concrete example in Appendix X.X" and show it in details. (3) We better choose less than 5 reasons, and map each of them into one of the designed components/features in the framework. It is not good if we propose many challenges, but our design doesn't link back to these challenges. }

% We further manually identify why the models fail in these tasks, and conclude the challenges as below:

%% file: Tex/5-Methodology.tex
\section{\tool{}}\label{sec:methodology}

\subsection{Design Overview}
To address the challenges mentioned above, we propose a novel solution named \tool{}. Our approach is inspired by the operational framework of real-world IR teams, specifically the Blue Team~\cite{blueteam}. In these teams, leaders leverage a comprehensive understanding of the infrastructure to formulate strategies, decompose them into manageable sub-tasks, and delegate these to individual executors, orchestrating the entire IR effort. The executors independently carry out their assigned tasks and provide feedback to the leader. The leader evaluates the outcomes, reflects on identified issues, and refines their decisions, thereby iteratively advancing IR.
% Our design is inspired by working mechanisms of real-world IR teams. 

Based on this workflow, we design \tool{} with three dynamic phases—Reasoning, Action, and Reflection—and coordinate its operation through four LLM session components: Planner, Generator, Reflector, and Analyst, as illustrated in Figure~\ref{fig:workflow}. The Planner formulates the strategic overall plan and decomposes it into manageable sub-tasks; the Generator produces specific instructions or commands based on Planner’s directives; the Reflector provides optimization suggestions by reflecting on various types of information; and the Analyst thoroughly analyzes the execution results and feeds them back to Planner. Each phase maintains an independent LLM session to preserve its specific session history and contextual information. Each component performs its designated role without interfering with others, thereby enhancing the efficiency and generalization capabilities of \tool{}. This structured division of labor, rooted in real-world practices, enables \tool{} to mitigate LLM limitations and achieve greater precision.
% the Reflector engages in reflection and provides optimization suggestions on various types of information; 
% This clearly defined division of labor, inspired by real-world practices, helps LLMs overcome their inherent limitations to a certain extent and achieve higher accuracy.

Drawing lessons from failures, we incorporate several innovations into the design of \tool{} to enhance its practicality and effectiveness. 
(1) To bolster the reasoning capabilities of LLMs, we employ a range of prompting methodologies (\ie, CoT~\cite{wei2022chain}, ToT~\cite{yao2024tree}, Few-Shot Prompt~\cite{min2022rethinking}, and Negative Prompt~\cite{rombach2022high}). 
(2) We refine the responsibilities of each phase, enabling the LLM to deliver more precise decisions while minimizing the risk of overlooking critical details. 
(3) To address Challenges I and IV, we introduce an IR Tree (\Sref{sec:methodology:Planner}), which integrates two key elements: an IR objective table and detailed IR steps.
(4) We design the Reflector component (\Sref{sec:methodology:Reflector}) to reflect on four distinct sources of information and offer adaptive optimization suggestions for improved decision-making. 
The complete prompts are available at our anonymous GitHub~\cite{IRCopilot}. Detailed design specifics for each component are provided below.

%---------------------------------------------------------------------------------
\subsection{Planner}\label{sec:methodology:Planner}

To overcome the challenges of strategy formulation and the loss of session context, we develop the Planner component. The Planner serves as the core decision-making component in \tool{}, responsible for overseeing IR tasks at a macro level, making critical decisions, and coordinating the entire IR process. The Planner receives IR outcomes or intents from the user and prepares for subsequent analysis and actions. These steps are then forwarded to the Generator for further detailed interpretation and execution.

\begin{figure}[!t]
  \centering
  \includegraphics[width=\linewidth]{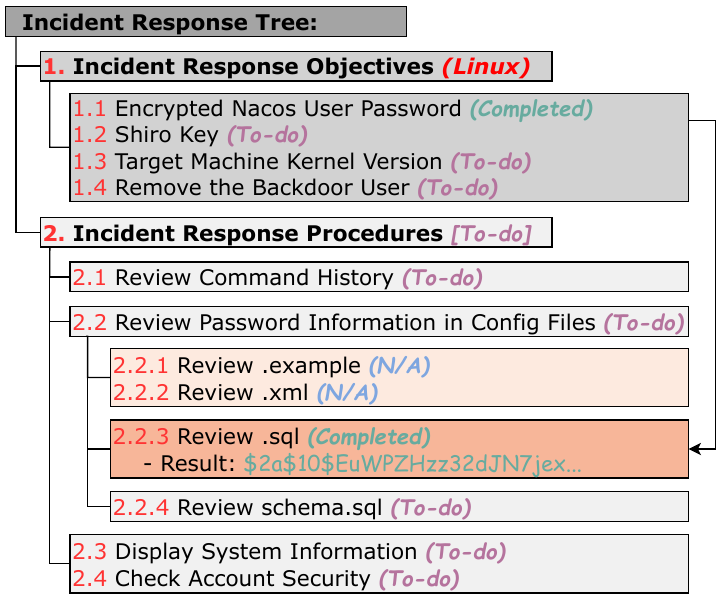}
  % \vspace{-2em}
  \caption{Natural Language Representation of the IRT for the XuanJi-Nacos Target.}
  \label{fig:IRT-NL}
  \vspace{-1em}
\end{figure}

IR is inherently continuous and multifaceted, demanding the formulation of long-term strategies and the preservation of contextual coherence. We draw lessons from Failure Case 1 and inspiration from the Penetration Testing Task Tree~\cite{deng2024pentestgpt}, motivating the design of \tool{} and leading to the introduction of the IRT (Incident Response Tree) as a core mechanism for task decomposition and tracking. The IRT organizes tasks in a hierarchical tree and assigns each node one of three statuses (\ie, \textit{Completed, To-do, and N/A}) to clarify objectives at each phase and maintain end-to-end continuity. By formalizing status tracking, the IRT effectively mitigates the contextual drift often encountered by LLMs during prolonged IR engagements.

Taking the XuanJi-Nacos target as an example, we demonstrate the operational mechanism of the Planner and its corresponding prompt strategy (\Fref{fig:prompt}), which we divide into reasoning and decision-making processes.

% The IRT tracks task statuses (\ie, "Completed," "To-do," and "N/A") in a structured tree format. This structured approach enhances task clarity at each stage and ensures the continuity of the IR process, thereby mitigating context loss in LLMs.

% The operation of the Planner encompasses the following key steps, with Figure~\ref{fig:IRT-NL} illustrating the natural language representation of the IRT for the XuanJi-Nacos target. 

% --------------------------------------------------------------------------------------------------------------------------
\begin{figure}[]
\begin{codebox}
As the leader of cybersecurity incident response, you are responsible for high-level planning and maintaining an Incident Response Tree (IRT). Adhere to the following principles:\\
\textbf{1. Task Structure:}
\begin{itemize}[leftmargin=1.5em,noitemsep,topsep=0pt]
\item  Organize all tasks in a hierarchical sequence (e.g., 1, 1.1, 1.1.1), where sub-tasks are nested under their parent tasks to form a clear tree-like structure.
\item  Each level reflects the dependency and granularity of tasks, ensuring logical progression from high-level objectives to specific actions.
\end{itemize}
% \\
\textbf{2. Task Status and Updates:}
\begin{itemize}[leftmargin=1.5em,noitemsep,topsep=0pt]
\item  Assign each task a status: "To Do," "Completed," or "N/A," and update it based on the latest findings with a brief outcome report.
\item  For tasks under "1. Incident Response Objectives," replace the status with specific details (e.g., answers or information) in parentheses once resolved (e.g., "1.1 OS version - (Ubuntu 20.04)"). Note: "N/A" is not permitted in this section.
\item  For other sections (e.g., "2. Incident Response Procedures"), retain the standard status labels and append results separately when applicable.
\end{itemize}
% \\
\textbf{3. Adding Sub-Tasks:}
\begin{itemize}[leftmargin=1.5em,noitemsep,topsep=0pt]
\item  Only add sub-tasks if a task is unclear or requires further investigation (e.g., analyzing historical command outputs). Avoid including unverified or undiscovered information in the IRT.
\end{itemize}
% \\
\textbf{4. Prohibited Commands:}
\begin{itemize}[leftmargin=1.5em,noitemsep,topsep=0pt]
\item Avoid global search commands like "find" or "grep."
\end{itemize}
% \\
Below are the IRT templates generated under two different scenarios:

\end{codebox}
\vspace{-0.8em}
% --------------------------------------------------------------------------------------------------------------------------
\begin{examplebox}
\textbf{Scenario 1}: When tasks are clear (e.g., specific information is required via a defined method), focus solely on "1. Incident Response Objectives" without additional sections:\\
% \\
\textbf{1. Incident Response Objectives (linux) - [To-do]}\\
\hspace*{1em}1.1 Server OS version - (To-do)\\
\hspace*{1em}1.2 Sensitive files in home directory - (To-do)\\
\hspace*{1em}...
\end{examplebox}
\vspace{-0.8em}
% --------------------------------------------------------------------------------------------------------------------------
\begin{mybox}
\textbf{Scenario 2}: When tasks lack clarity (e.g., involving a "flag field"), expand the IRT with "2. Incident Response Procedures" for further investigation:\\
% \\
\textbf{1. Incident Response Objectives (linux) - [To-do]}\\
\hspace*{1em}1.1 Attacker IP - (...)\\
\hspace*{1em}1.2 Modified plaintext admin password - (To-do)\\
\hspace*{1em}...\\
\textbf{2. Incident Response Procedures - [To-do]}\\
\hspace*{1em}2.1 Review Command History - (Completed)\\
\hspace*{2em}Results from 2.1: - ...\\
\hspace*{1em}2.2 Investigate Sensitive Directories - (To-do)\\
\hspace*{1em}...
% \hspace*{1em}2.3 Analyze System Logs - (To-do)\\
% \hspace*{1em}2.4 Check Open Ports and Services - (To-do)\\
% \hspace*{2em}Results from 2.4:\\
% \hspace*{2em}- Port 21: ftp\\
% \hspace*{2em}- Port 22: ssh\\
% \hspace*{1em}2.5 Investigate Abnormal Behavior - (To-do)\\
% \hspace*{2em}2.5.1 Investigate Processes - (To-do)\\
% \hspace*{2em}2.5.2 Investigate Startup Items - (To-do)\\
% \hspace*{2em}2.5.3 Investigate Cron Jobs - (To-do)\\
% \hspace*{1em}2.6 Check Account Security - (To-do)\\
% \hspace*{2em}Results from 2.6:\\
% \hspace*{2em}- Suspicious account: ...\\
% \hspace*{1em}2.7 Investigate Suspicious Files - (To-do)\\
% \hspace*{2em}2.7.1 Check Modified Files - (To-do)\\
% \hspace*{1em}2.8 Check Database Content - (To-do)\\
\end{mybox}
\vspace{-1em}
\caption{The prompt strategy of Planner for two scenarios: principles in the initial part, followed by specific strategies for Scenario 1 (clear tasks) and Scenario 2 (unclear tasks).}
\label{fig:prompt}
\end{figure}
% --------------------------------------------------------------------------------------------------------------------------

\noindent\textbf{(1) Initialization and Classification of the IRT \ding{182}:} 
The Planner acquires the user's IR objectives and fundamental system information to construct the initial IRT structure, represented as a hierarchical task table in natural language. We conduct a scenario-based categorization and analysis of user intentions. When clear IR goals are provided, \tool{} initiates the response using these as the starting point, and logging them in the "\textit{Incident Response Objectives}" table (see Section 1 in~\Fref{fig:IRT-NL}). For scenarios lacking explicit objectives, we assign priorities to IR sub-tasks based on expert insights from~\Sref{sec:benchmark}, and maintain a secondary "\textit{Incident Response Procedures}" tree under the root node (see Section 2 in~\Fref{fig:IRT-NL}). \tool{} begins the investigation with high-priority tasks, progressively addressing subsequent ones. Throughout this process, the Planner classifies and labels tasks. Simultaneously, it performs a preliminary deduction of potential execution steps within the task table, laying the foundation for subsequent IR processes.
% By systematically maintaining response objectives and tasks, it mitigates the risk of context loss that may occur in LLMs.

\noindent\textbf{(2) Verification and Maintenance of the IRT \ding{183}:} 
After the IRT is generated, the Reflector verifies it to ensure that irrelevant tasks are excluded, preventing unnecessary objectives from being mistakenly incorporated into the IRT. During the verification process, the Reflector prevents modifications to the root node or key IR objectives of the IRT, thereby avoiding structural deviations or errors in response steps caused by LLM hallucinations. If any issues are detected during verification, the Reflector triggers a rollback mechanism, prompting the Planner to re-evaluate and deduce the IRT to ensure task accuracy and correctness.

\noindent\textbf{(3) Dynamic Sub-task Management \ding{184}:} 
Following the update of the IRT, the Planner prioritizes recently added sub-tasks, incorporating them into the pending task list and formulating concise solutions by weighing various contextual factors. In Figure~\ref{fig:IRT-NL}, this is reflected by the prioritization of the recently added \textit{2.2.x sub-tasks}. Specifically, the sub-task \textit{"2.2.3 Review nacos-mysql.sql"} is completed, while tasks such as \textit{"2.2.4 Review schema.sql,"} \textit{"2.1 Review Command History,"} and \textit{"2.3 Display System Information"} remain in the To-do status. \ding{185} The Planner will evaluate these tasks, determine their priority, and formulate decisions to refine the execution path. Subsequently, the tasks are delegated to the Generator for step-by-step analysis and generation of execution instructions.

\noindent\textbf{(4) Continuous Optimization and Feedback Loop \ding{186}:} 
To optimize the continuous update process of the IRT, we introduce a result recording mechanism at IRT nodes (e.g., \textit{Result: \$2a\$10\$...}). This ensures that task execution statuses and outputs are updated in real-time within the IRT, combating the forgetting issue associated with LLMs. Additionally, the system conducts real-time reflection on decisions, and the Planner dynamically adjusts IRT branches based on these results, continuously optimizing subsequent task paths to improve IR efficiency.

Furthermore, for ethical and security concerns, we design an interface that enables direct interaction with the Planner to prevent LLM from unintentionally accessing sensitive data.

%---------------------------------------------------------------------------------
\subsection{Generator}
To address the challenges of inaccurate guidance or commands, we designed the Generator. The Generator plays a critical role in guiding and generating the sub-tasks determined by the Planner, responsible for converting abstract task requirements into concrete executable commands. Upon receiving a new sub-task, \tool{} directs the Generator to disregard extraneous contextual information, focus exclusively on the current sub-task, and avoid interference from other sub-tasks. 

In this process, we do not have the Generator produce guidance directly; instead, it decomposes the task into several sequential steps. \ding{187} Firstly, the Generator focuses on refining the sub-tasks formulated by the Planner into a series of detailed steps. 
% Based on the present circumstances, and particularly considering the different operating systems specified in the IRT incident response objectives, the Generator is instructed to develop specific guidance strategies. 
We briefly tag the OS in \textit{IRT-Incident Response Objectives} and prompt the Generator to formulate specific strategies for different OS based on the current context. \ding{188} When a sub-task can be accomplished in multiple ways, the Generator creates as many distinct guidance strategies as possible. Subsequently, it converts these guidance strategies into executable terminal commands or provides precise investigative descriptions within desktop operating systems. \ding{189} Throughout this process, the Generator not only ensures that the semantics and logic of the commands align with expectations but also strictly standardizes their format by using the "\$" symbol as both the starting and ending markers. This formatting approach facilitates subsequent parsing and extraction of key information while minimizing parsing ambiguities caused by inconsistent formats, thereby enhancing the system's robustness and the accuracy of task execution.

The Generator serves as a pivotal component connecting the forward-looking decisions provided by the Planner with the specific execution steps of IR. It guarantees that strategic plans are carefully converted into accurate and executable operational instructions. This conversion process transforms abstract strategic intentions into concrete practical actions, thereby significantly enhancing the coherence and overall efficiency of IR execution. Furthermore, it produces clear and human-friendly outputs, ensuring the integrity and traceability of the response process.

%---------------------------------------------------------------------------------
\subsection{Reflector}\label{sec:methodology:Reflector}
The Reflector is capable of effectively analyzing and reflecting on the natural language interactions occurring among the other three central components. The rationale of introducing this component is primarily based on the following four needs. First, it mitigates privacy and security concerns, as LLMs may unintentionally access sensitive data, posing significant risks to data security during task execution. Second, contemporary LLMs still suffer from hallucination issues, which may cause other components to understand or generate erroneous information, thereby directly leading to the failure of IR tasks or increasing unnecessary steps and costs. Third, users who lack specialized IR knowledge find it difficult to determine whether the LLM has made suboptimal decisions, generated incorrect guidance, or neglected key information. Finally, in iterative scenarios, the Reflector prevents the accumulation of errors by monitoring and correcting outputs, ensuring other components do not perpetuate initial mistakes when handling new tasks or expanded contexts. 
Thus, the Reflector is essential for resolving these challenges.

In \tool{}, the Reflector processes four distinct types of information: (1) \ding{183} IRT consists of content generated by the Planner and updated by the Analyst; (2) \ding{186} decisions made by the Planner together with concise descriptions; (3) \ding{190} guidance or commands produced by the Generator based on the Planner’s decisions and descriptions; and (4) \ding{191} execution results of the Generator’s guidance. Drawing inspiration from methodologies such as Few-Shot Prompt and Negative Prompt, we develop a set of prompts that give representative error examples and their possible causes and empower the Reflector to identify potential hallucinations across contexts, confirm the inclusion of all critical information, pinpoint erroneous steps and their causes, and ultimately suggest modifications and conduct traceback analysis.

%---------------------------------------------------------------------------------
\subsection{Analyst}
%Penetration testing~\cite{weidman2014penetration} typically aims to gain access to a single target and follows a cyclical, iterative process. 
IR requires the comprehensive analysis of multiple distinct traces left by attackers, such as intrusion paths, malware activities, indications of data breaches, and anomalous behaviors in system logs. This process takes the form of a one-to-many, potentially weighted, tree structure. Therefore, our IRT design must comprehensively consider the possibilities of various tasks and their inter-dependencies. Additionally, the issue of overlooking minor but crucial details must be taken into account in the design.

To address these issues, we design the Analyst as an auxiliary module to the Planner. It adopts the Tree-of-Thought (ToT)~\cite{yao2024tree} reasoning technique, which generalizes over CoT to prompting language models, and enables exploration over coherent units of thoughts that serve as intermediate steps toward problem solving. ToT enables LLMs to engage in intentional decision-making by evaluating various reasoning trajectories and reflecting on their choices to determine the subsequent actions. Additionally, ToT facilitates forward planning and retrospective adjustments when required, ensuring that decisions are made with a comprehensive and global perspective. The Analyst breaks down the complete results into multiple executable sub-tasks. In particular, once the execution results for specific sub-tasks are received, it dynamically adds items requiring further investigation to the IRT. Finally, the Analyst feeds these findings back to the Planner, helping it make more precise and effective decisions.

%% file: Tex/6-Evaluation.tex
\section{Evaluation on \tool{}}\label{sec:evaluation}
In this section, we assess \tool{} by addressing the following research questions:

\begin{itemize}[leftmargin=*,noitemsep,topsep=0pt]
% \item \textbf{RQ1 (Performance \& Efficiency):} How does \tool{} perform with different LLMs in terms of performance and efficiency?
\item \textbf{RQ1 (Performance):} How does \tool{} perform with different underlying LLMs in IR?

\item \textbf{RQ2 (Efficiency):} How efficient is \tool{} in terms of response time and monetary cost?

\item \textbf{RQ3 (Ablation):} What is the contribution of each component within \tool{} to the overall IR?

\item \textbf{RQ4 (Practicality):} Does \tool{} demonstrate practicality and effectiveness in real-world IR tasks?

\item \textbf{RQ5 (Failure):} Under what conditions does \tool{} fail, and what are the underlying causes of these failures?

% \item \textbf{RQ4 (Efficiency):} Is \tool{} efficient in real-time IR?

\end{itemize}

\subsection{Experimental Settings}
We implement \tool{} with 2,788 lines of Python3 code and 257 lines of prompts. We evaluate its performance by utilizing the benchmark tests outlined in \Sref{sec:benchmark} and applying it to real-world IR scenarios. We develop 6 working versions, \toolgptfour{}, \toolllama{}, \tooldeepseek{}, \toolgptfouro{}, \toolclaude{}, and \toolgptoone{}, using \gptfour{}~\cite{API}, \llama{}~\cite{dubey2024llama}, \deepseek{}~\cite{deepseekAPI} \gptfouro{}~\cite{API}, \claude{}~\cite{claudeAPI} and \gpto{}~\cite{API} as the underlying LLMs respectively. Except for the prompts we designed, all other inputs and experimental settings are identical to those described in \Sref{sec:benchmark:evaluation}.

%------------------------------------------------------------------------------------------------------------------------------------
\subsection{Performance Evaluation (RQ1)}\label{sec:evaluation:PE}

\begin{figure}[]
  \centering
  \begin{subfigure}{\linewidth}
    \centering
    \includegraphics[width=\linewidth]{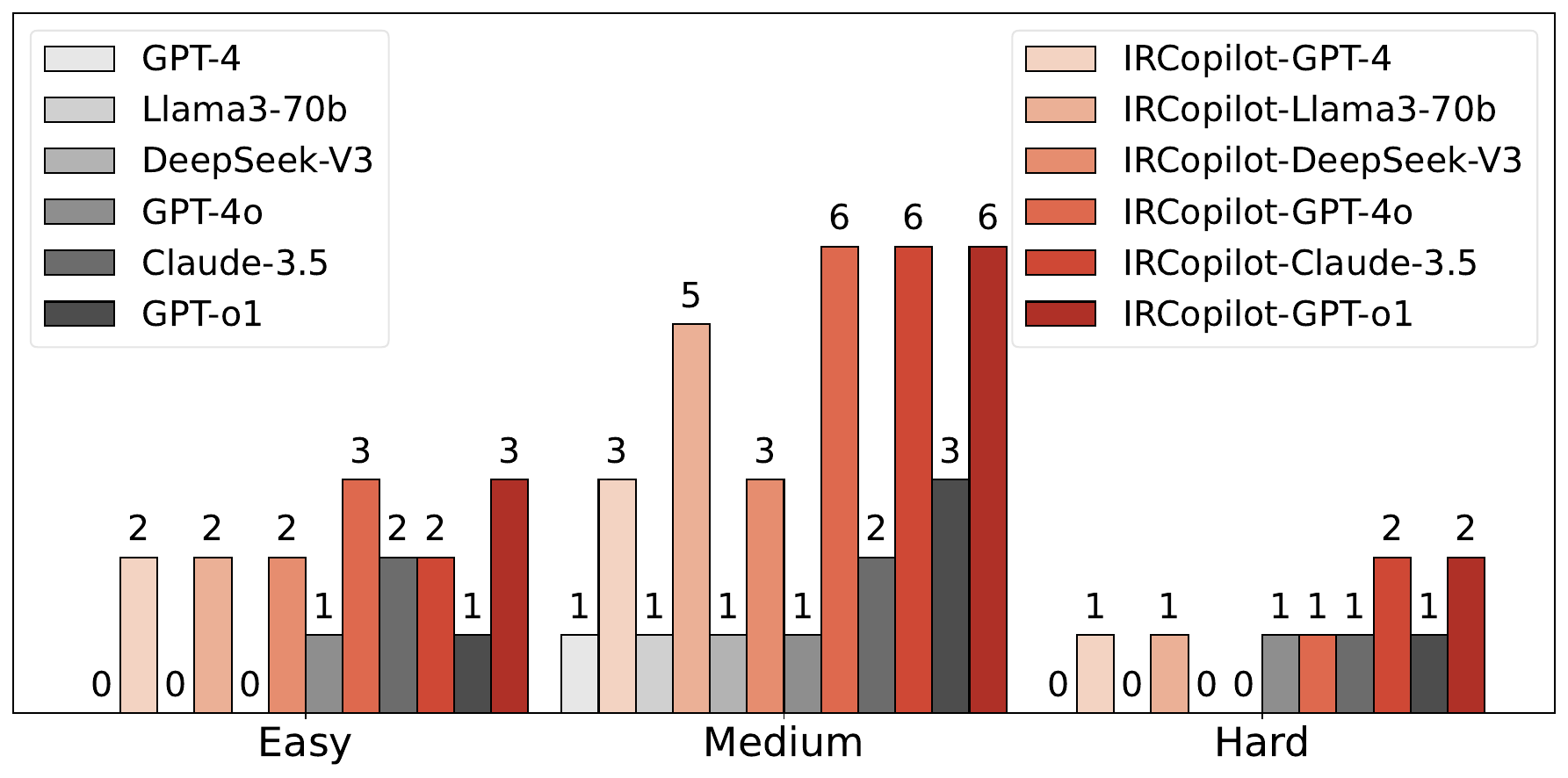}
    % \vspace{-2em}
    \caption{Overall completion status.}
	  \label{fig:task}
  \end{subfigure}
  
  \begin{subfigure}{\linewidth}
    \centering
    \includegraphics[width=\linewidth]{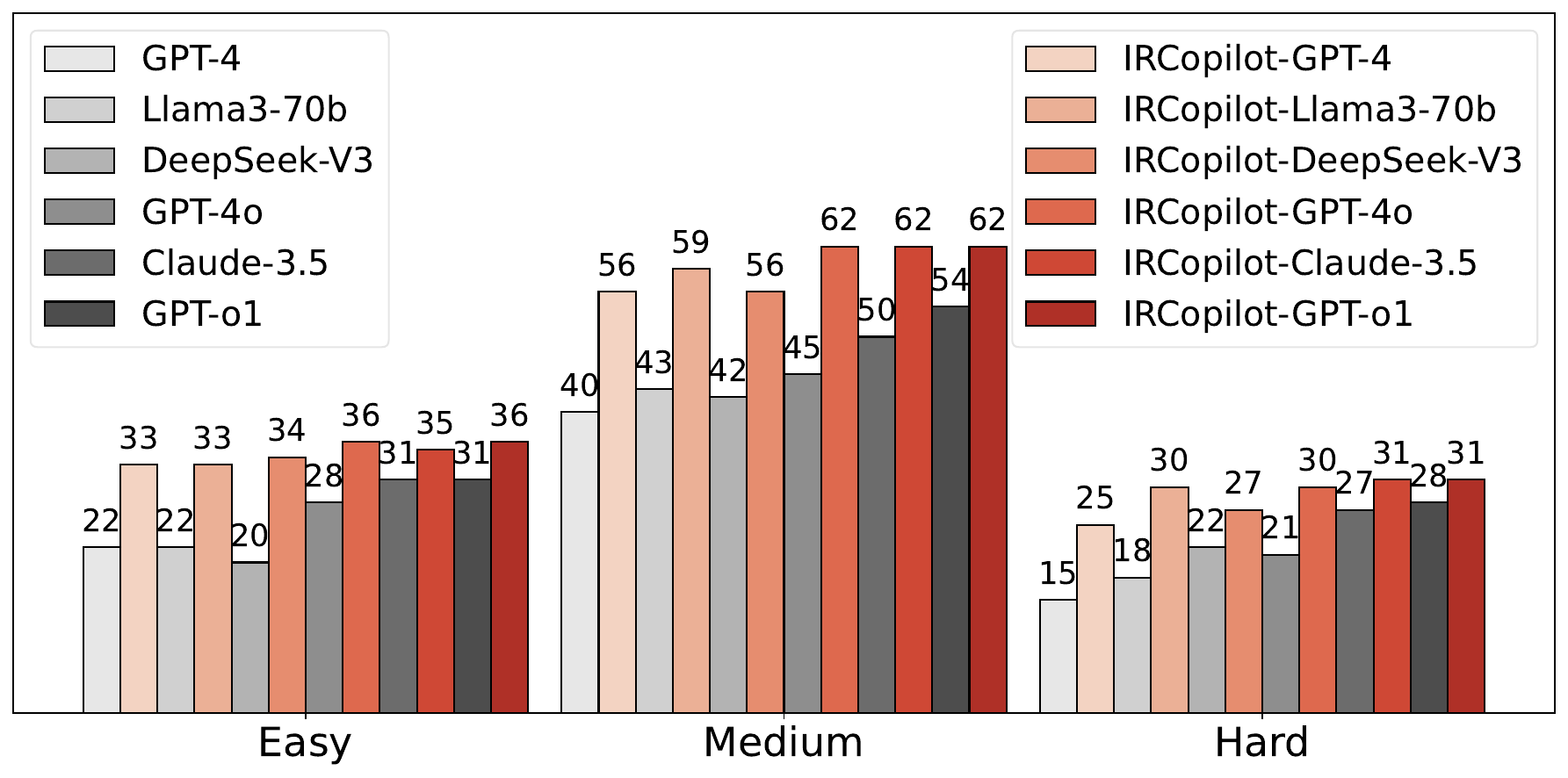}
    % \vspace{-2em}
    \caption{Sub-task completion status.}
	  \label{fig:subtask}
  \end{subfigure}
  % \vspace{-1em}
  \caption{The performance of \gptfour{}, \toolgptfour{}, \llama{}, \toolllama{}, \deepseek{}, \tooldeepseek{}, \gptfouro{}, \toolgptfouro{}, \claude{}, \toolclaude{}, \gpto{}, and \toolgptoone{} on IRBench.}
  \label{fig:performance}
  % \vspace{-1em}
\end{figure}

\Fref{fig:task} presents the performance of different LLMs and \tool{} counterparts across tasks of varying difficulty. As shown, \tool{} significantly boosts the performance of its base LLMs in IR tasks. In particular, \toolgptfouro{}, \toolclaude{}, and \toolgptoone{} outperform other models at most difficulty levels, exhibiting a notable advantage in handling incidents of moderate or lower complexity. In contrast, \toolgptfour{} and \gptfour{} both trail behind \toolgptfouro{} and \gptfouro{}, revealing the limitations of the \gptfour{}-based approach in more complex scenarios. We attribute this discrepancy to the relatively smaller training dataset and narrower incident coverage of \gptfour{} compared with \gptfouro{}, a limitation also noted by OpenAI~\cite{API}.

\Fref{fig:subtask} further details the performance at the sub-task level. As illustrated, \toolgptfour{}, \toolllama{}, \tooldeepseek{}, \toolgptfouro{}, \toolclaude{}, and \toolgptoone{} each outperform their respective base LLMs, achieving 150\%, 147\%, 138\%, 136\%, 119\%, and 114\% of the base-model performance, respectively. This outcome confirms that our approach effectively alleviates critical bottlenecks faced by LLMs in IR workflows, such as hallucination, context loss, and difficulties with complex, progressive questioning. This further verifies the applicability and effectiveness of our approach. However, these methods still encounter difficulties with high-complexity tasks, primarily due to inherent limitations of LLMs that remain unaddressed. Our approach does not expand the training knowledge base in domains such as attacks, investigations, analysis, or response, which limits improvements in more complex scenarios, as further discussed in~\Sref{sec:discussion}.

In summary, \tool{} substantially improves the IR performance of all tested LLMs, demonstrating its effectiveness as a framework. Having established this performance gain, we next evaluate the associated efficiency costs.

%------------------------------------------------------------------------------------------------------------------------------------
\subsection{Efficiency Evaluation (RQ2)}\label{sec:evaluation:efficiency}

% We calculate the average time per sub-task by dividing the total time by the number of completed sub-tasks for a more precise efficiency evaluation. The metrics are defined as follows. 
% (1) $n_g$ and $n_i$: Number of sub-tasks completed by \gptfouro{} and \tool{}. 
% (2) $\overline{t}_g$ and $\overline{t}_i$: Average time to complete each task for \gptfouro{} and \tool{} (in seconds). 
% (3) $\overline{s}_g$ and $\overline{s}_i$: Average time per sub-task for \gptfouro{} and \tool{} (in seconds), computed as $\overline{t}_g / n_g$ and $\overline{t}_i / n_i$ to reflect efficiency.

To answer RQ2, we analyze the efficiency of \tool{} by measuring its response time and monetary cost, thereby quantifying the trade-off between its enhanced performance and the resources required. We conduct a study on IRBench under strictly controlled conditions. We ensure consistency by performing all experiments at the same time and under identical environmental settings, with other configurations matching those described in previous sections. Likewise, we utilize the \gptfouropre{} API to integrate \toolgptfouropre{} seamlessly. For each task in the benchmark, we perform ten experiments to measure the time taken by LLM to generate response content.

Real‑world IR is subject to unpredictable factors such as network delays and varying incident complexity. To mitigate this, we focus on the LLM's reasoning time, defined as the duration to process inputs and produce outputs on platform, as a critical and controllable metric. This approach isolates the LLM's performance, emphasizing its reasoning speed as the key efficiency metric. Our goal is to provide reference indicators for efficiency in real-time IR systems. 
% \Tref{tab:efficiency} compares the efficiency of \gptfouro{} and \tool{} on benchmark. 

To formalize this, we define $t_{j,k}$ as the reasoning time of method $j$ (where $j=g$ for \gptfouro{} and $j=i$ for \tool{}) in the $k$‑th trial. The average task completion time over $K$ runs is:

\begin{equation}
\overline{t}_j = \frac{1}{K} \sum_{k=1}^{K} t_{j,k}, \quad j \in \{g, i\},
\label{eq:avg_task_time}
\end{equation}

The core purpose of our efficiency experiment is not just to measure total time, but to understand how efficiently each method solves problems. Because \tool{} completes more sub-tasks ($n_i > n_g$), a direct comparison of total time ($\overline{t}_j$) is misleading. We, therefore, introduce a normalized metric: the average time per completed sub-task, which better reflects problem-solving efficiency:

\begin{equation}
\overline{s}_j = \frac{\overline{t}_j}{n_j}, \quad j \in \{g, i\},
\label{eq:avg_subtask_time}
\end{equation}

where $n_g$ and $n_i$ are the counts of sub-tasks completed by \gptfouro{} and \tool{}, respectively.

% \textcolor{red}{Our method generally requires more time across most tasks, owing to the overhead of multi‑step reasoning and maintenance of the IRT. Nonetheless, it outperforms \gptfouro{} on four of the twelve benchmark tasks and we consider this trade-off of efficiency for enhanced performance to be justified, as evidenced by a 15\% accuracy improvement in 'Ransomware' despite a 20\% time increase. We hope that future research can improve efficiency while maintaining performance. Importantly, even with this additional time, our approach remains more efficient than traditional manual IR methods~\cite{GreyTime, MeanTime}.}
\Tref{tab:efficiency} shows that while \tool{} often requires more total time ($\overline{t}_i$) due to the overhead of multi-step reasoning, it is more efficient on a per-sub-task basis ($\overline{s}_i$) in four of the twelve benchmarks. This indicates that although the overall process is longer, \tool{} is more direct and effective in solving individual steps. For instance, in the 'Ransomware' task, \tool{} achieves a 50\% increase in completed sub-tasks ($n_i=6$ vs. $n_g=4$) for only a 35\% increase in total time. This trade-off—sacrificing some speed for significantly better performance—is highly valuable in critical IR scenarios. Moreover, even with this overhead, our approach remains far more efficient than traditional manual IR methods~\cite{GreyTime, MeanTime}.

\begin{table}[t]
	\centering
	\caption{Efficiency comparison of \gptfouropre{} and \toolgptfouropre{} on IRBench.}
	\fontsize{14}{16}\selectfont
	% \vspace{-0.5em}
	\resizebox{\linewidth}{!}{
		\begin{tabular}{l| c c| c c| c c}
			\toprule
			\textbf{Benchmark}    & \textbf{$n_g$} & \textbf{$n_i$} & \textbf{$\overline{t}_g$ (s)} & \textbf{$\overline{t}_i$ (s)} & \textbf{$\overline{s}_g$ (s)} & \textbf{$\overline{s}_i$ (s)} \\ \midrule
			Investigating Windows & 17             & 22             & \textbf{249.9}                & 278.0                         & 14.7                          & \textbf{12.6} \\
			Linux 1               & 4              & 7              & \textbf{61.6}                 & 193.6                         & \textbf{15.4}                 & 27.7 \\
			Web 1                 & 7              & 7              & \textbf{69.7}                 & 147.4                         & \textbf{10.0}                 & 21.1 \\
			Tardigrade            & 12             & 15             & \textbf{143.7}                & 201.0                         & \textbf{12.0}                 & 13.4 \\
			Ransomware            & 4              & 6              & \textbf{118.0}                & 159.2                         & 29.5                          & \textbf{26.5} \\
			Web 2                 & 7              & 11             & \textbf{54.3}                 & 109.1                         & \textbf{7.8}                  & 9.9 \\
			Web 3                 & 7              & 11             & \textbf{83.3}                 & 289.3                         & \textbf{11.9}                 & 26.3 \\
			Black Pages \& Tampering & 10          & 10             & \textbf{62.0}                 & 112.8                         & \textbf{6.2}                  & 11.3 \\
			Windows Miner         & 5              & 9              & 92.0                          & \textbf{68.1}                 & 18.4                          & \textbf{7.6} \\
			Linux 2               & 7              & 10             & \textbf{66.1}                 & 294.0                         & \textbf{9.4}                  & 29.4 \\
			Nacos                 & 6              & 6              & \textbf{75.9}                 & 119.4                         & \textbf{12.7}                 & 19.9 \\
			Where-1S-tHe-Hacker   & 8              & 14             & \textbf{108.2}                & 157.1                         & 13.5                          & \textbf{11.2} \\ \bottomrule
		\end{tabular}
	}
	\label{tab:efficiency}
	% \vspace{-1em}
\end{table}

\begin{table}[t]
	\centering
	\caption{Comparison of \toolgptfouropre{} and \toolgptoone{} on IRBench.}
	\fontsize{14}{16}\selectfont
	% \vspace{-0.5em}
	\resizebox{\linewidth}{!}{
		\begin{tabular}{l| c c| c c| c c}
			\toprule
			\textbf{Benchmark}    & \textbf{$n_{4o}$} & \textbf{$n_{o1}$} & \textbf{$\overline{t}_{4o}$ (s)} & \textbf{$\overline{t}_{o1}$ (s)} & \textbf{$\overline{c}_{4o}$ (USD)} & \textbf{$\overline{c}_{o1}$ (USD)} \\ \midrule
			Investigating Windows & 22                & 22                & \textbf{278.0}                   & 914.4                            & \textbf{0.78}                      & 5.90 \\
			Linux 1               & 7                 & 7                 & \textbf{193.6}                   & 463.8                            & \textbf{0.27}                      & 2.06 \\
			Web 1                 & 7                 & 7                 & \textbf{147.4}                   & 1,095.5                          & \textbf{0.45}                      & 3.79 \\
			Tardigrade            & 15                & 15                & \textbf{201.0}                   & 673.3                            & \textbf{0.43}                      & 3.54 \\
			Ransomware            & 6                 & 6                 & \textbf{159.2}                   & 664.0                            & \textbf{0.23}                      & 2.77 \\
			Web 2                 & 11                & 11                & \textbf{109.1}                   & 708.7                            & \textbf{0.43}                      & 3.45 \\
			Web 3                 & 11                & 11                & \textbf{289.3 }                  & 1,365.8                          & \textbf{0.65}                      & 6.48 \\
			Black Pages \& Tampering & 10             & 10                & \textbf{112.8}                   & 674.5                            & \textbf{0.46}                      & 2.17 \\
			Windows Miner         & 9                 & 9                 & \textbf{68.1}                    & 577.5                            & \textbf{0.35}                      & 1.94 \\
			Linux 2               & 10                & \textbf{11}       & \textbf{294.0}                   & 885.7                            & \textbf{0.51}                          & 4.73 \\
			Nacos                 & 6                 & 6                 & \textbf{119.4}                   & 630.0                            & \textbf{0.32}                      & 2.81 \\
			Where-1S-tHe-Hacker   & 14                & 14                & \textbf{157.1}                   & 690.7                            & \textbf{0.35}                          & 3.88 \\ \bottomrule
		\end{tabular}
	}   
	\label{tab:reason}
	% \vspace{-1em}
\end{table}
%------------------------------------------------------------------------------------------------------------------------------------

% In \Tref{tab:reason}, we further compare \tool{} with a reasoning‑focused LLM (\gpto{}) against \tool{} with \gptfouropre{} on IRBench. Here, $n_{4o}$ and $n_{o1}$ denote sub‑task counts; $\overline{t}_{4o}$ and $\overline{t}_{o1}$ the average running times; and $\overline{c}_{4o}$ and $\overline{c}_{o1}$ the corresponding costs.
% (1) $n_{4o}$ and $n_{o1}$: Number of sub-tasks completed by \toolgptfouropre{} and \toolgptoone{}. 
% (2) $\overline{t}_{4o}$ and $\overline{t}_{o1}$: Average time to complete each task for \toolgptfouropre{} and \toolgptoone{} (in seconds). 
% (3) $\overline{c}_{4o}$ and $\overline{c}_{o1}$: Average cost to complete each task for \toolgptfouropre{} and \toolgptoone{} (USD).
% \textcolor{red}{In \Tref{tab:reason}, we further compare \tool{} with a reasoning‑focused LLM (\gpto{}) against \tool{} with \gptfouropre{} on IRBench. Three metrics are reported:}
% (1) Task coverage ($n_{4o}$, $n_{o1}$): number of sub-tasks completed by \toolgptfouropre{} and \toolgptoone{}, respectively.
% (2) Average reasoning time ($\overline{t}_{4o}$, $\overline{t}_{o1}$): mean time per task in seconds for each configuration.
% (3) Average monetary cost ($\overline{c}_{4o}$, $\overline{c}_{o1}$): mean cost per task in USD for each configuration.

Finally, we analyze the monetary cost. We compare \toolgptfouropre{} with \toolgptoone{} on IRBench, reporting task coverage ($n_{4o}$, $n_{o1}$), average reasoning time ($\overline{t}_{4o}$, $\overline{t}_{o1}$), and average cost per task ($\overline{c}_{4o}$, $\overline{c}_{o1}$). The cost is formalized analogously to time. Let $c_{j,k}$ be the cost incurred by method $j$ in the $k$-th trial. Over $K$ runs, the average cost per task is:

\begin{equation}
\overline{c}_j = \frac{1}{K} \sum_{k=1}^{K} c_{j,k}, \quad j \in \{4o, o1\},
\label{eq:avg_cost_task}
\end{equation}

% The outcomes demonstrate that \tool{} exhibits versatility across LLMs with different thinking patterns. However, the outcomes also reveal that these patterns lead to varying performance characteristics. In particular, the use of \gpto{} yields a modest performance improvement, but this enhancement is accompanied by a substantial increase in reasoning time and cost overhead. 
% \textcolor{red}{The results show that \tool{} is adaptable to LLMs with distinct reasoning styles, yet those styles lead to different efficiency profiles. Using \gpto{} delivers a modest accuracy gain, but at the expense of markedly higher reasoning time and cost. Overall, \tool{} paired with \gptfouro{} currently offers the best overall trade-off among performance, speed, and cost for real-time IR.}
The results show that \tool{} is adaptable to LLMs with distinct reasoning styles, yet those styles lead to different efficiency profiles. While \toolgptoone{} delivers a marginal performance gain (e.g., completing one extra sub-task in 'Linux 2'), it comes at the expense of significantly higher reasoning time and monetary cost. In summary, \tool{} paired with \gptfouro{} currently offers the best trade-off among performance, speed, and cost for real-time IR applications.

%------------------------------------------------------------------------------------------------------------------------------------
\subsection{Ablation Study (RQ3)}
To analyze the contribution of each component in our approach to the overall IR process, we conduct an ablation study, design and implement the following four variants:
\begin{enumerate}[leftmargin=*,noitemsep,topsep=0pt]
\item \toolnoplanner{}: the Planner component is disabled, bypassing the IRT strategy and forwarding tasks directly to subsequent components.
\item \toolnogenerator{}: the Generator component is deactivated, allowing others to directly produce guidance.
\item \toolnoreflector{}: the Reflector component is disabled, eliminating the reflection step.
\item \toolnoanalyst{}: the Analyst component and the ToT prompting method are disabled, directing execution results straight to the Planner component.
\end{enumerate}

\toolgptfouro{} and \toolclaude{} demonstrate similar performance in \Sref{sec:evaluation:PE}. However, \toolgptfouro{} stands out with significantly lower costs, approximately 1/5 of \toolclaude{}’s, along with faster response times. Due to its superior cost-effectiveness and efficiency, we consider \toolgptfouro{} the preferred choice. Consequently, all variants utilize the \toolgptfouro{} API for evaluation. Notably, a comprehensive IR task encompasses multiple objectives, and success requires fulfilling all of them. Conducting ablation experiments on the components could therefore hinder successful task completion. For this reason, our ablation study focuses solely on the quantitative analysis of sub-tasks.

Figure~\ref{fig:ablation-performance} illustrates the performance of the four variants assessed through our comprehensive benchmark tests.  Across these evaluations, \tool{} consistently surpasses each ablation baseline in sub-task completion rates. The ablation study yields several key observations, summarized as follows:

\begin{figure}[t]
  \centering
    \includegraphics[width=\linewidth]{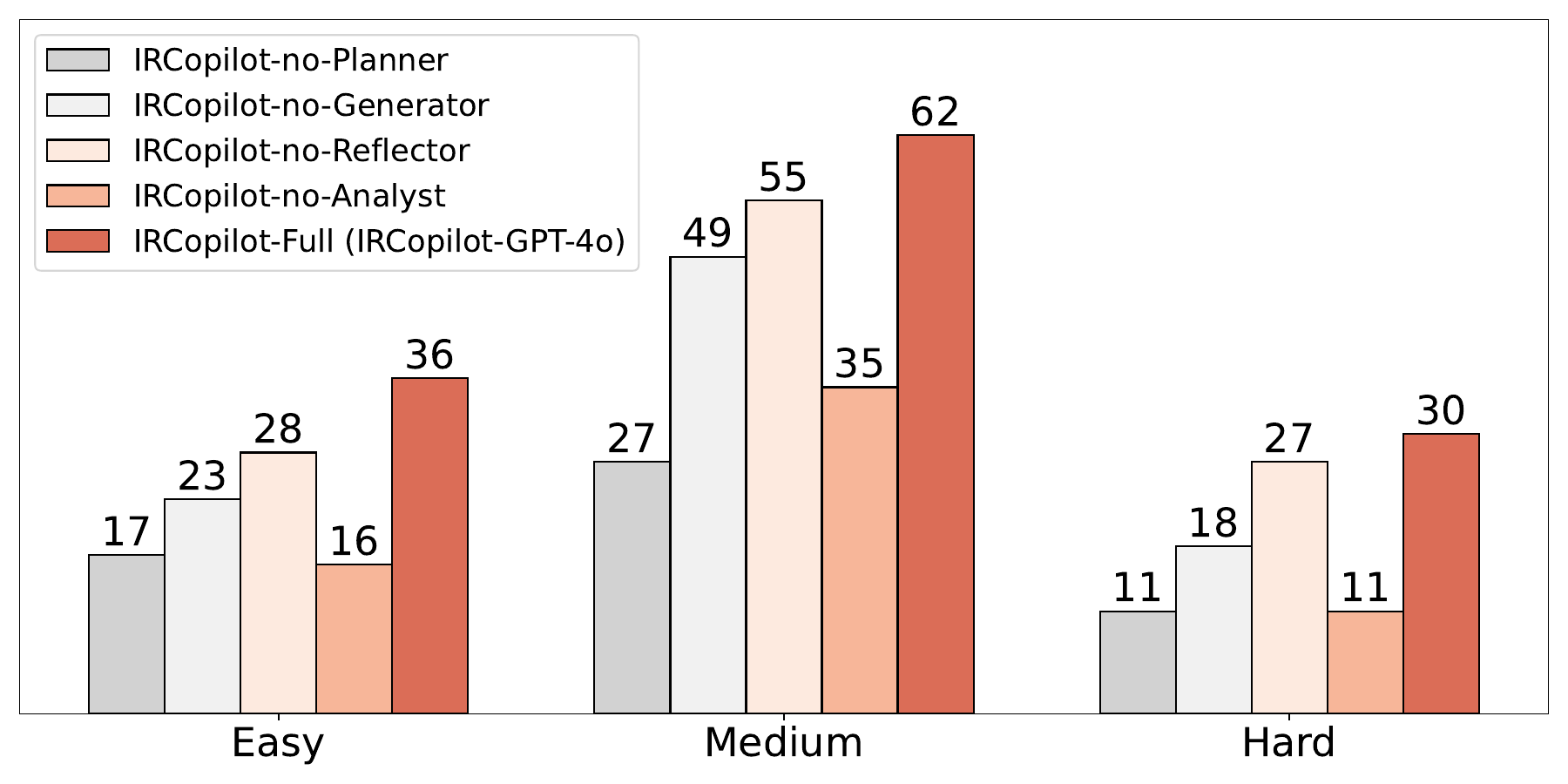}
    % \vspace{-2em}
    \caption{The performance of \tool{}, \toolnoplanner{}, \toolnogenerator{}, \toolnoreflector{} and \toolnoanalyst{} on IRBench.}
  \label{fig:ablation-performance}
  % \vspace{-1em}
\end{figure}

\begin{itemize}[leftmargin=*,noitemsep,topsep=0pt]
\item \toolnoplanner{} achieves the lowest success rate, completing only 42.97\% (55/128) of the sub-tasks compared to the full variant, followed by \toolnoanalyst{} at 48.44\% (62/128). These figures fall below even the baseline \gptfouro{}’s performance. This highlights the critical importance of the Planner component’s IRT strategy and decision-making capabilities, as well as the Analyst component’s ToT method, in ensuring effective IR processes. Moreover, these results emphasize that robust reasoning ability is indispensable for driving successful outcomes in such tasks.

\item \toolnogenerator{} performs approximately on par with the baseline \gptfouro{} at 70.31\% (90/128), indicating that this component primarily generates IR guidance through the base LLM. This provides auxiliary support to incident responders and reduces operational costs.

\item \toolnoreflector{} demonstrates a slight decrease compared to the full variant in sub-task completion at 85.94\% (90/128). This decline is mainly attributed to occasional LLM 'hallucinations' or other minor issues. However, the frequency of such errors is relatively low, and the Reflector component is capable of rectifying these minor errors in most cases, thereby positively contributing to the overall IR process. Nevertheless, \toolnoreflector{} still marginally outperforms the baseline \gptfouro{}, suggesting that even in the absence of the Reflector component, the other components of \tool{} continue to provide a certain performance advantage.
\end{itemize}

%------------------------------------------------------------------------------------------------------------------------------------
% \vspace{-1em}
\subsection{Practicality Study (RQ4)}\label{sec:evaluation:practicality}
To evaluate the adaptability of \tool{} in addressing real-world cybersecurity incidents, we conducted a practicality study that extends beyond the benchmark tests outlined in \Sref{sec:benchmark}. For this assessment, we selected 35 distinct machines across three different platforms to test \tool{}'s performance:

\begin{enumerate}[leftmargin=*,noitemsep,topsep=0pt]
\item TryHackMe~\cite{TryHackMe}: A subscription-based cybersecurity education platform. We selected 17 machines of varying difficulty: 7 easy, 9 medium, and 1 hard target.
\item XuanJi~\cite{XuanJi}: A pay-per-time platform featuring globally recognized IR machines. We select 12 representative machines, consisting of 2 easy, 6 medium, and 4 hard targets.
\item ZGSF~\cite{ZGSF}: A public account platform managed by a cybersecurity laboratory, providing free access. From this, we select 6 representative machines, including 2 easy, 3 medium, and 1 hard target. These machines are adapted from real-world cyber attack cases, with relevant privacy information removed to prevent ethical and moral issues.
\end{enumerate}

Our evaluation integrates \tool{} with the \gptfouropre{} API to create \toolgptfouropre{}, establishing the completion of the entire IR process as the criterion for a successful test. For each target, we conduct five experiments and record the number of times the tasks are successfully completed. \textit{We establish that if at least one of the five experiments results in a successful response, we consider the IR successful.} This is because in real-world cybersecurity IR process, multiple security practitioners usually work individually on one particular workload to perform extended analyses and responses. Therefore, the ultimate success contingent upon the successful completion of the task from at least one of the security experts. 

\begin{table}[!t]
	\centering
	\caption{\tool{} performance over TryHackMe challenges.}
	\fontsize{14}{16}\selectfont
	\resizebox{\linewidth}{!}{
		\begin{tabular}{l|c c c c}
			\hline
			\textbf{Challenge}         & \textbf{Difficulty} & \textbf{OS}       & \textbf{Completions}      & \textbf{Cost (USD)} \\ \hline
			Eradication \& Remediation & easy                & Linux             & 4/5 (\ding{51})           & 0.29 \\
			Investigating Windows      & easy                & Windows           & 4/5 (\ding{51})           & 0.78 \\
			Linux Incident Surface     & easy                & Linux             & 3/5 (\ding{51})           & 0.53 \\
			Linux File System Analysis & easy                & Linux             & 5/5 (\ding{51})           & 0.46 \\
			Linux Process Analysis     & easy                & Linux             & 5/5 (\ding{51})           & 0.69 \\
			Threat Intel \& Containment& easy                & Linux             & 5/5 (\ding{51})           & 0.08 \\
			Windows Network Analysis   & easy                & Windows           & 5/5 (\ding{51})           & 0.15 \\
			Blizzard                   & medium              & Windows           & 0/5 (\ding{55})           & - \\
			Investigating Windows 2.0  & medium              & Windows           & 4/5 (\ding{51})           & 1.11 \\
			Investigating Windows 3.x  & medium              & Windows           & 4/5 (\ding{51})           & 0.80 \\
			Linux Logs Investigations  & medium              & Linux             & 5/5 (\ding{51})           & 0.29 \\
			Linux Live Analysis        & medium              & Linux             & 4/5 (\ding{51})           & 0.54 \\
			Tardigrade                 & medium              & Linux             & 3/5 (\ding{51})           & 0.43 \\
			Tempest                    & medium              & Windows           & 0/5 (\ding{55})           & - \\
			Windows Applications Forensics& medium           & Windows           & 4/5 (\ding{51})           & 0.31 \\
			Windows Event Logs         & medium              & Windows           & 5/5 (\ding{51})           & 0.24 \\
			Squid Game                 & hard                & Linux             & 0/5 (\ding{55})           & - \\ \hline
		\end{tabular}
	}
	\label{tab:TryHackMe-challenges}
	% \vspace{-1em}
\end{table}

\begin{table}[!t]
	\centering
	\caption{\tool{} performance over XuanJi challenges.}
	\fontsize{14}{16}\selectfont
	\resizebox{\linewidth}{!}{
		\begin{tabular}{l|c c c c}
			\hline
			\textbf{Challenge}                & \textbf{Difficulty}& \textbf{OS}       & \textbf{Completions}      & \textbf{Cost (USD)} \\ \hline
			Webshell Detection and Removal    & easy               & Linux             & 4/5 (\ding{51})           & 0.24 \\
			Windows EVTX File Analysis        & easy               & Windows           & 5/5 (\ding{51})           & 0.21 \\
			Linux Backdoor Emergency          & medium             & Linux             & 2/5 (\ding{51})           & 0.80 \\
			Log Analysis - MySQL              & medium             & Linux             & 0/5 (\ding{55})           & - \\
			Log Analysis - Redis              & medium             & Linux             & 0/5 (\ding{55})           & - \\
			Traffic Feature Analysis - Tomcat & medium             & Linux             & 3/5 (\ding{51})           & 0.24 \\
			VulnTarget n - Ransomware         & medium             & Linux             & 4/5 (\ding{51})           & 0.23 \\
			Windows Black Pages \& Tampering & medium             & Windows           & 5/5 (\ding{51})           & 0.46 \\
			Memory Trojan Analysis - Nacos    & hard               & Linux             & 5/5 (\ding{51})           & 0.32 \\
			Memory Trojan Analysis - shiro    & hard               & Linux             & 5/5 (\ding{51})           & 0.19 \\
			Windows Wordpress                 & hard               & Windows           & 3/5 (\ding{51})           & 0.26 \\
			Where 1S tHe Hacker               & hard               & Windows           & 0/5 (\ding{55})           & - \\ \hline
		\end{tabular}
	}
	\label{tab:XuanJi-challenges}
	% \vspace{-1em}
\end{table}

% \begin{table}[h]
% \caption{\tool{} performance over ZGSF challenges.}
% \resizebox{\linewidth}{!}{
% \begin{tabular}{l|c c c c}
% \hline
% \textbf{Challenge}    & \textbf{Difficulty}& \textbf{OS}       & \textbf{Completions}      & \textbf{Cost (USD)} \\ \hline
% Linux 1               & easy               & Linux             & 5/5 (\ding{51})           &0.27        \\ 
% Windows 1             & easy               & Windows           & 5/5 (\ding{51})           &0.45        \\ 
% Windows 2             & medium             & Windows           & 3/5 (\ding{51})           &0.43        \\
% Windows 3             & medium             & Windows           & 4/5 (\ding{51})           &0.65        \\ 
% Cryptojacking         & medium             & Windows           & 5/5 (\ding{51})           &0.35        \\ 
% Linux 2               & hard               & Linux             & 0/5 (\ding{55})           &-           \\ \hline
% \end{tabular}
% }
% \label{tab:ZGSF-challenges}
% % \vspace{-1em}
% \end{table}

\begin{table}[!t]
	\centering
	\caption{\tool{} performance over ZGSF challenges.}
	\fontsize{14}{16}\selectfont
	\resizebox{\linewidth}{!}{
		\begin{tabular}{l|c c c c}
			\hline
			\textbf{Challenge}    & \textbf{Difficulty}& \textbf{OS}       & \textbf{Completions}      & \textbf{Cost (USD)} \\ \hline
			Linux 1               & easy               & Linux             & 5/5 (\ding{51})           &0.27        \\ 
			Windows 1             & easy               & Windows           & 5/5 (\ding{51})           &0.45        \\ 
			Windows 2             & medium             & Windows           & 3/5 (\ding{51})           &0.43        \\
			Windows 3             & medium             & Windows           & 4/5 (\ding{51})           &0.65        \\ 
			Cryptojacking         & medium             & Windows           & 5/5 (\ding{51})           &0.35        \\ 
			Linux 2               & hard               & Linux             & 0/5 (\ding{55})           &-           \\ \hline
		\end{tabular}
	}
	\label{tab:ZGSF-challenges}
	% \vspace{-1em}
\end{table}

\Tref{tab:TryHackMe-challenges} details the performance of \tool{} in the TryHackMe challenges. \tool{} successfully completes all 7 easy tasks and 7 of 9 medium tasks, achieving a completion rate of 77.8\%. Concurrently, when utilizing the \gptfouropre{} API, \tool{} maintains a low operational cost. This indicates that \tool{} is capable of efficiently handling a wide range of low to medium difficulty of IR tasks with reduced expenditure. Notably, the platform offers few high-difficulty tasks, and only one is available, which \tool{} does not complete.

\Tref{tab:XuanJi-challenges} outlines the performance of \tool{} on the XuanJi platform challenges. At the easy level, \tool{} completes all tasks, demonstrating high reliability. For medium challenges, it resolves 3 of 6 tasks (50\% completion rate) at an average cost of \$0.39 per task. At the hard level, it completes 2 of 4 challenges (50\% completion rate), showcasing its capability for complex tasks. In general, \tool{} exhibits strong adaptability, although performance in higher-difficulty tasks could be improved.

\Tref{tab:ZGSF-challenges} presents \tool{}'s performance on ZGSF challenges. It achieves a 100\% completion rate for both easy and medium tasks, with average costs of \$0.36 and \$0.48 per task, respectively. However, \tool{} fails to complete the hard-level tasks, indicating certain inherent limitations in handling highly complex IR scenarios.

Notably, we included fewer tasks of Hard difficulty. This is due to the limited number of publicly available IR tasks on the platform. Despite this constraint, we ensured that the selected tasks were sufficient to support our conclusions. Specifically, compared to the other two difficulty levels, IRCopilot exhibits slightly lower performance on Hard difficulty tasks.

% \Tref{tab:ZGSF-challenges} illustrates the detailed performance of \tool{} in the ZGSF platform challenges. At the easy level, \tool{} again achieves a 100\% completion rate, with an average cost of approximately \$0.36 per task. Similarly, at the medium level, \tool{} consistently maintains a 100\% completion rate, with an average cost of \$0.48 per task. However, in the hard challenges, \tool{} is unable to complete the tasks, indicating certain inherent limitations in handling highly complex IR scenarios.

Additionally, we select five real-world instances of personal PCs or servers compromised by attacks, which are published on the XuanJi platform. These cases span diverse system types, attack scenarios, and techniques. Similarly, to adhere to ethical guidelines, the privacy data of these machines has been anonymized.

% \begin{table}[t]
% \caption{Performance over real-world attack cases.}
% \resizebox{\linewidth}{!}{
% \begin{tabular}{l|c c c c}
% \hline
% \textbf{Machine}    & \textbf{OS}      & \textbf{Tasks} & \textbf{Completions}      & \textbf{Cost (USD)} \\ \hline
% Sunlogin            & Windows          & 3              & 4/5 (\ding{51})           & 0.66       \\ 
% algo                & Linux            & 4              & 3/5 (\ding{51})           & 0.51       \\ 
% kswapd0             & Linux            & 3              & 5/5 (\ding{51})           & 0.25       \\ 
% Cryptojacking 1     & Linux            & 5              & 2/5 (\ding{51})           & 0.90       \\
% Cryptojacking 2     & Linux            & 6              & 4/5 (\ding{51})           & 0.61       \\ \hline
% \end{tabular}
% }
% \label{tab:case-challenges}
% % \vspace{-1em}
% \end{table}

\begin{table}[!t]
	\centering
	\caption{Performance over real-world attack cases.}
	\fontsize{14}{16}\selectfont
	\resizebox{\linewidth}{!}{
		\begin{tabular}{l|c c c c}
			\hline
			\textbf{Machine}    & \textbf{OS}      & \textbf{Tasks} & \textbf{Completions}      & \textbf{Cost (USD)} \\ \hline
			Sunlogin            & Windows          & 3              & 4/5 (\ding{51})           & 0.66       \\ 
			algo                & Linux            & 4              & 3/5 (\ding{51})           & 0.51       \\ 
			kswapd0             & Linux            & 3              & 5/5 (\ding{51})           & 0.25       \\ 
			Cryptojacking 1     & Linux            & 5              & 2/5 (\ding{51})           & 0.90       \\
			Cryptojacking 2     & Linux            & 6              & 4/5 (\ding{51})           & 0.61       \\ \hline
		\end{tabular}
	}
	\label{tab:case-challenges}
	% \vspace{-1em}
\end{table}

\Tref{tab:case-challenges} showcases capabilities of \tool{} in adapting to tasks on compromised machines within real-world attack environments, demonstrating its effectiveness in accurately identifying and extracting attacker traces and artifacts from genuine cyber incidents. This capability is crucial for significantly improving the efficiency of handling and responding to cybersecurity incidents. The findings indicate that \tool{} operates robustly and demonstrates considerable potential across a variety of real and diverse challenges, providing valuable support for post‑IR and bolstering defenses against the evolving cyber threat landscape.

%------------------------------------------------------------------------------------------------------------------------------------
\subsection{Failure Analysis (RQ5)}\label{sec:evaluation:failure}

\begin{table}[!t]
  \centering
  \caption{Failure Reasons of \tool{} Combined with Different LLMs Variants on IRBench}
  \vspace{-0.5em}
  % 在大括号里先切换字体尺寸，再做宽度缩放
  {%
    \fontsize{14}{16}\selectfont  % 字体 12pt，行距 14pt，可根据需要调整
    \resizebox{\linewidth}{!}{%
      \begin{tabular}{l|*{6}{c}|c}
		\toprule
		\textbf{Failure Reasons}       & \textbf{GPT4} & \textbf{DeepSeek} & \textbf{Llama}  & \textbf{4o}  & \textbf{Claude}   & \textbf{o1}  & \textbf{Total}\\ \midrule
		Key Information Ignored        & 5             & 6              & 4                  & 1            & 0                & 0                & \textbf{16}\\ 
		False Command Generation       & 5             & 1              & 1                  & 1            & 1                & 1                & \textbf{10}\\
		False Guidance Generation      & 4             & 3              & 2                  & 0            & 0                & 0                & \textbf{9}\\
		False IR Strategy              & 2             & 3              & 1                  & 0            & 1                & 0                & \textbf{7}\\ \midrule
		Total                          & 16            & 13             & 8                  & 2            & 2                & 1               & \textbf{42}\\ \bottomrule
      \end{tabular}%
    }%
  }%
  \label{tab:IR-failure}
  % \vspace{-2em}
\end{table}

We conduct a failure analysis of \tool{} across the sub-tasks of IRBench, detailed in~\Tref{tab:IR-failure}. The analysis reveals that the application of \tool{} significantly reduces the failure rate from 28.3\% to 5.4\%, and this improvement is particularly pronounced for more foundational models. More importantly, we observe a significant shift in the distribution of primary failure modes. The dominant error transitions from "False IR Strategy"—a common issue when using LLMs directly—to "Key Information Ignored" and "False Command Generation." We attribute this shift to a critical trade-off:

\begin{itemize}[leftmargin=*,noitemsep,topsep=0pt]
\item While \tool{}'s structured approach mitigates strategic failures, its maintenance of state and multi-turn interactions increase the volume of information the LLM must process. This heightened cognitive load can cause the model to overlook critical details, thus explaining the rise in "Key Information Ignored" errors.
    
\item The LLM's inherent limitations, namely its static knowledge base and susceptibility to hallucinations, can lead to the invention of incorrect commands or guidance, accounting for "False Command Generation."
\end{itemize}    

To complement these quantitative findings, we perform a fine-grained qualitative analysis on the challenges \tool{} failed in~\Sref{sec:evaluation:practicality}, moving beyond a binary verdict of success or failure to understand the nuances of the agent's reasoning process. This analysis is structured as a cross-validation between human experts and Gemini 2.5 Pro~\cite{gemini}—an independent, third-party LLM not utilized in our preceding experiments—to ensure an objective assessment of the intermediate logical steps. This in-depth analysis explores the root causes of failure, identifying fundamental deficiencies in the reasoning process such as "Hallucinations" and "Limitations in Reverse," which we discuss further in~\Sref{sec:discussion}. To illustrate this evaluation process, the Appendices~\ref{appendix:case3} and~\ref{appendix:case4} present summaries of the most critical parts of two failed case studies. The complete, step-by-step evaluations for these two cases are available in our anonymized GitHub repository. Ultimately, this focus on reasoning quality offers critical insights for enhancing the reliability and trustworthiness of automated incident response workflows.

%% file: Tex/7-Discussion.tex
\section{Discussion}\label{sec:discussion}

\textbf{Data Contamination Mitigation.} 
\tool{} is built on LLMs, recognizing that cases or write‑ups on these platforms may already be included in LLM training data, which can potentially bias experimental results. To address concerns about potential data contamination in LLMs and \tool{}, we implement three measures. Initially, we prioritize benchmark cases with more recent publication dates during the selection process, which are most likely to fall outside the scope of the LLMs' training data. We then compare the publication date of each selected case with the specific knowledge cutoff dates of the LLMs~\cite{cutoff, openaicutoff} as shown in~\Tref{tab:Cutoff}. Majorities (43/72) of the tasks occur after the cutoff date of the models. Subsequently, we assess whether the LLMs underwent targeted training by querying them for detailed information about specific cases. More details are available at our Anonymous Github~\cite{IRCopilot}. 
% for transparency
Finally, since \tool{} has the ability to display each step of its reasoning process, we confirm through an examination of its reasoning paths that it lacks prior knowledge of the cases. Furthermore, real-world cases in practicality study further demonstrated that, even without targeted training, \tool{} shows significant feasibility and effectiveness in real-world applications.
% , thus minimizing the impact of pre-training data

% \textcolor{blue}{
% \textbf{Difficulty Selection.} 
% In the experiments of benchmark and \Sref{sec:evaluation:practicality}, we included fewer tasks of Hard difficulty. This is due to the limited number of publicly available IR tasks on the platform. Despite this constraint, we ensured that the selected tasks were sufficient to support our conclusions. Specifically, compared to the other two difficulty levels, \tool{} exhibits slightly lower performance on Hard difficulty tasks.}

\textbf{Selection of LLM API.} 
During the design phase, we systematically evaluated multiple LLM APIs and ultimately select four: \gptfour{}, \gptfouropre{}, and \claude{} as commercial APIs, alongside \llama{} and \deepseek{} as open-source models. In our experiments, the latest model \gptfouropre{} shows even better performance than \gptfour{}, offering faster response and a context window of 128K tokens. Meanwhile, \claude{} and \deepseek{} excel at reasoning and code analysis. Notably, \gptfouropre{} also costs just 1/12 of \gptfour{}, making it a cost-effective choice for real-world applications. 
% While we initially consider the \gpto{} series, we ultimately exclude it from our study due to its lack of support for temperature parameter and slower reasoning speeds.

% \begin{table}[t]
% \caption{Comparison of IRBench release dates against LLM knowledge cutoff dates.}
% \resizebox{\linewidth}{!}{
% \begin{tabular}{l|c c c c c c}
% \toprule
% \textbf{Model}   & \textbf{GPT-4}  & \textbf{Llama3} & \textbf{DeepSeek-V3}  & \textbf{GPT-4o}  & \textbf{Claude-3.5}  & \textbf{GPT-o1} \\ \midrule
% \textbf{Cutoff}  & 2023-04              & 2023-12             & 2024-07               & 2023-10               & 2024-04             & 2023-10          \\ \midrule
% \textbf{After}   & 10                   & 10                  & 0                     & 10                    & 3                   & 10               \\ 
% \textbf{Before}  & 2                    & 2                   & 12                    & 2                     & 9                   & 2                \\ \bottomrule
% \end{tabular}
% }
% \label{tab:Cutoff}
% % \vspace{-1em}
% \end{table}

\begin{table}[!t]
	\centering
	\caption{Comparison of IRBench release dates against LLM knowledge cutoff dates.}
	\fontsize{14}{16}\selectfont
	\resizebox{\linewidth}{!}{
		\begin{tabular}{l|c c c c c c}
			\toprule
			\textbf{Model}   & \textbf{GPT-4}  & \textbf{Llama3} & \textbf{DeepSeek-V3}  & \textbf{GPT-4o}  & \textbf{Claude-3.5}  & \textbf{GPT-o1} \\ \midrule
			\textbf{Cutoff}  & 2023-04              & 2023-12             & 2024-07               & 2023-10               & 2024-04             & 2023-10          \\ \midrule
			\textbf{After}   & 10                   & 10                  & 0                     & 10                    & 3                   & 10               \\ 
			\textbf{Before}  & 2                    & 2                   & 12                    & 2                     & 9                   & 2                \\ \bottomrule
		\end{tabular}
	}
	\label{tab:Cutoff}
	\vspace{-1em}
\end{table}

\textbf{Limitations.} 
Even advanced models display limitations under specific conditions. For instance, they may produce outputs that contradict prior context, resulting in hallucinations~\cite{zhang2023siren, zhang2023language}. In our system, these hallucinations typically appear as erroneous analyses, judgments, or decisions derived from existing information (Appendices \ref{appendix:case3} and \ref{appendix:case4}). To address this, we introduce the IRT and Reflector to mitigate their impact. Beyond hallucinations, complex sample analysis and reverse engineering tasks often remain incomplete due to the limited capabilities of both \tool{} and LLMs in handling intricate scenarios (Appendix \ref{appendix:case5}). As training data and model architectures continue to evolve, we anticipate that LLM performance will improve, gradually reducing the occurrence of hallucinations and enhancing the ability to manage complex IR tasks.
% Although LLMs perform exceptionally well across numerous tasks, their capabilities remain constrained by multiple factors such as the scale of training data, model architecture, and specific application scenarios. Even advanced models exhibit limitations under certain conditions. For example, when outputs contradict previous context~\cite{zhang2023siren}, leading to "hallucinations"~\cite{zhang2023language}. In our system, these hallucinations primarily manifest as incorrect analyses, judgments, or decisions based on existing information, which can lead to failure of IR tasks. To mitigate their impact, we introduce the IRT and Reflector, combined with the ToT method. 
% As training data and model architectures evolve, we expect LLM performance to improve, reducing hallucinations over time.

% LLMs are prone to occasional hallucinations~\cite{zhang2023language}, which involve generating content that does not fully align with factual information or making non-optimal decisions based on available information. This issue somewhat compromises the robustness and consistency of our approach. To mitigate these intermittent hallucinations, we have developed the IRT and Reflector components and adopted the ToT methodology to minimize their negative impact on model performance. It is foreseeable that, over time, as training data and corpora continue to expand, and as model architectures and training algorithms are continuously optimized, the performance of LLMs will gradually improve, and the occurrence of hallucinations is expected to be mitigated to some extent.

\textbf{Future Work for Improvement.} 
Despite our efforts to design and evaluate multiple countermeasures, cybersecurity is always an evolving game, as we face the emergence of increasingly diverse and sophisticated attack vectors. We attempted to integrate RAG into the design, but it led to decreased efficiency and higher overhead, failing to yield the anticipated performance gains. Looking forward, a promising research direction is the development of adaptive defense mechanisms. These should incorporate more advanced agent systems, augmented with enhanced memory or search capabilities, to effectively counter sophisticated threats. Our findings indicate a positive correlation between the effectiveness of this approach and the capabilities of baseline LLMs. We are confident that, as LLMs evolve and defense strategies improve, these systems will become more adaptable, robust, and efficient, capable of addressing a broader spectrum of complex security incidents.

%% file: Tex/8-Conclusion.tex
\section{Conclusion}\label{sec:conclusion}

Modern cybersecurity landscape brings numerous challenges for automated incident response. This paper performs a systematic study to address them with two contributions. First, we establish a new comprehensive incident response benchmark to demonstrate the limitations of directly applying LLMs in this complex domain. Second, we design \tool{}, an advanced LLM-driven approach for efficient incident response by simulating the planning and response processes of real-world professional teams. This is achieved by an interactive modular design, and adoptions of multiple strategies to mitigate LLMs' hallucinations and context limitations. We evaluate \tool{} through extensive experiments and real-world scenarios, demonstrating the feasibility of leveraging LLMs for incident response tasks and laying the groundwork for future developments of cybersecurity enhancement.

%% file: Tex/Appendix.tex
\appendix

\section{Global Cybersecurity Landscape}\label{appendix:landscape}

\begin{figure}[!h]
  \centering
  \begin{subfigure}[]{\linewidth}
    \centering
    \includegraphics[width=\linewidth]{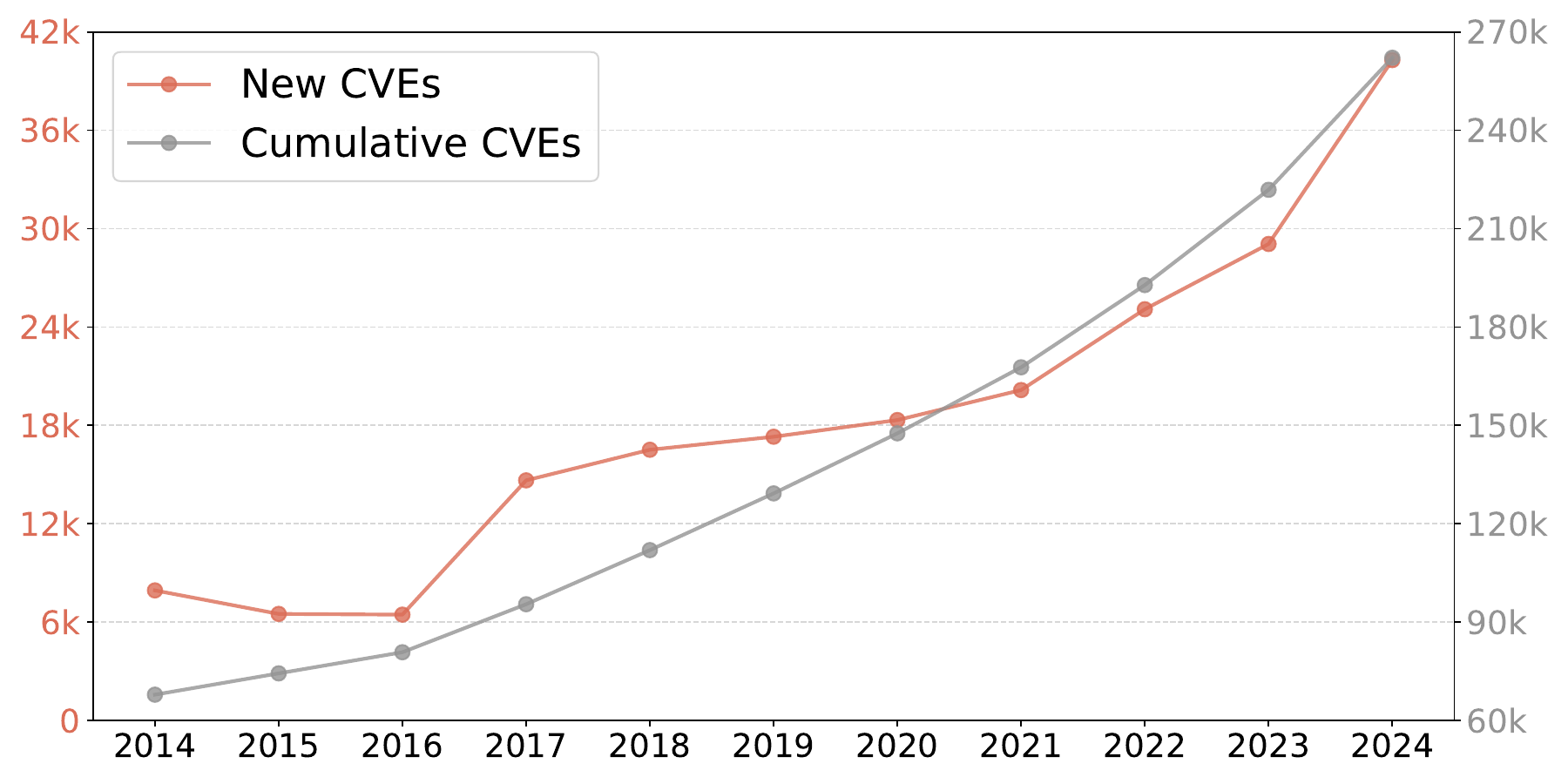}
    \vspace{-2em}
    \caption{CVE Trends (2014-2024).}
	\label{fig:CVE_Trends} 
  \end{subfigure}
  
  % Subfigure 3: Vulnerability Classification
  \begin{subfigure}{\linewidth}
    \centering

    \includegraphics[width=\linewidth]{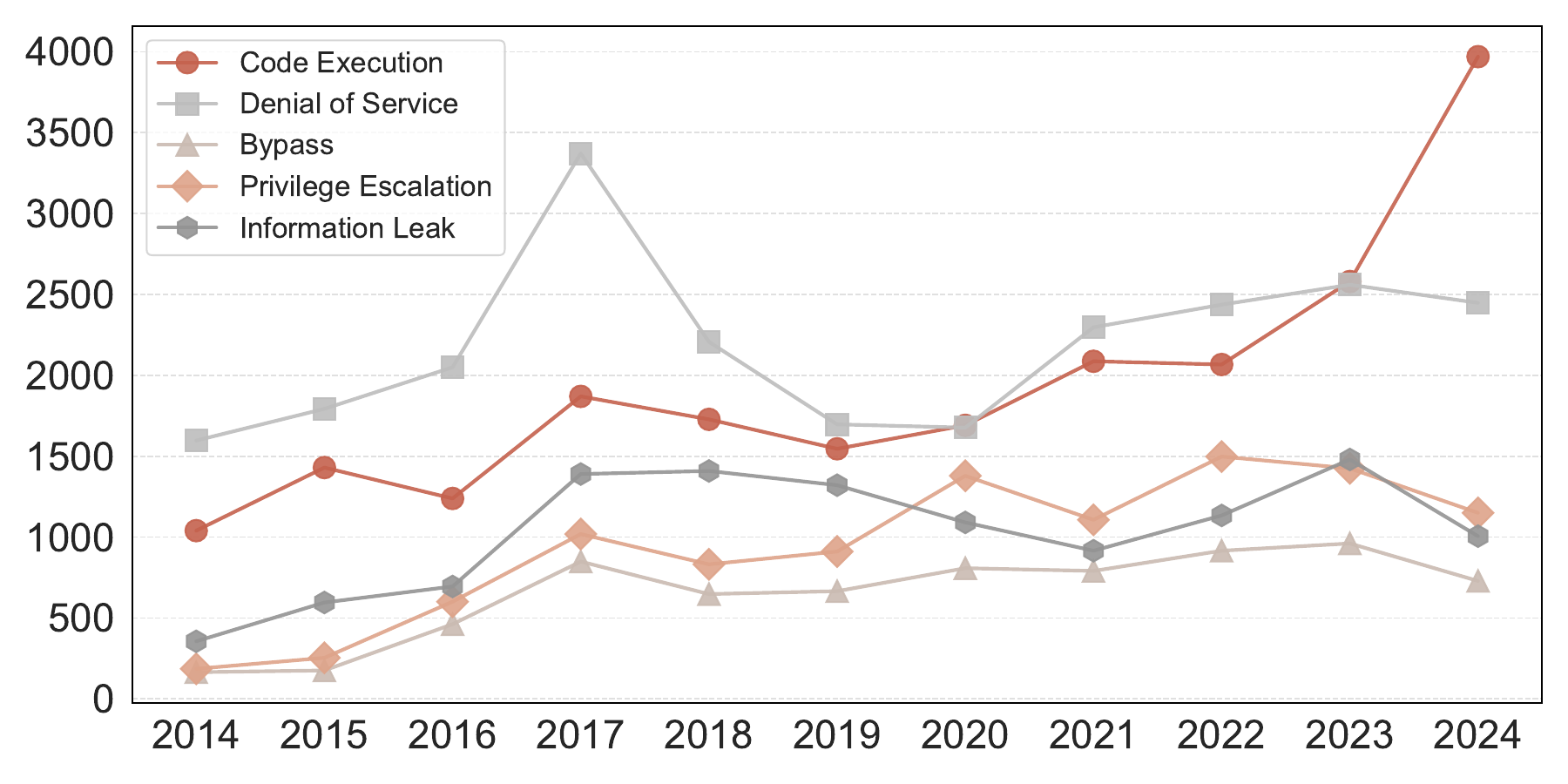}
    \vspace{-2em}
    \caption{Vulnerabilities by Impact Types (2014-2024).}
    \label{fig:Vulnerability-Classification}
  \end{subfigure}
  \vspace{-2em}
  \caption{Overview of CVEs (2014-2024).}
  \label{fig:CVEs}
  % \vspace{-1em}
\end{figure}

\begin{figure}[h]
  \centering
  \includegraphics[width=\linewidth]{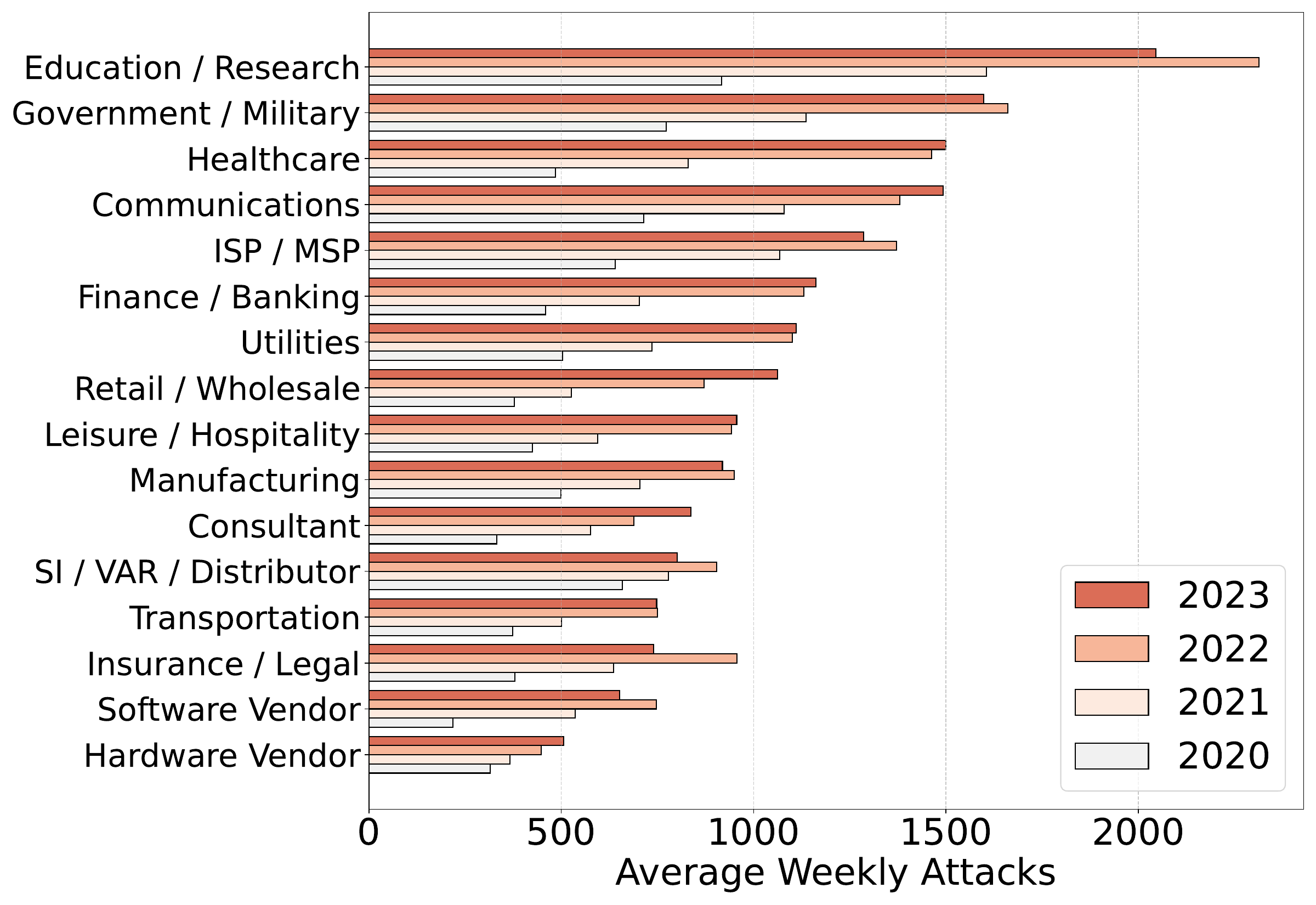}
  \vspace{-2em}
  \caption{Average Weekly Attacks per Organization by Industry (2020-2023).}
  \label{fig:Weekly-Attacks}
  \vspace{-2em}
\end{figure}

\begin{figure}[h]
  \centering
  \includegraphics[width=\linewidth]{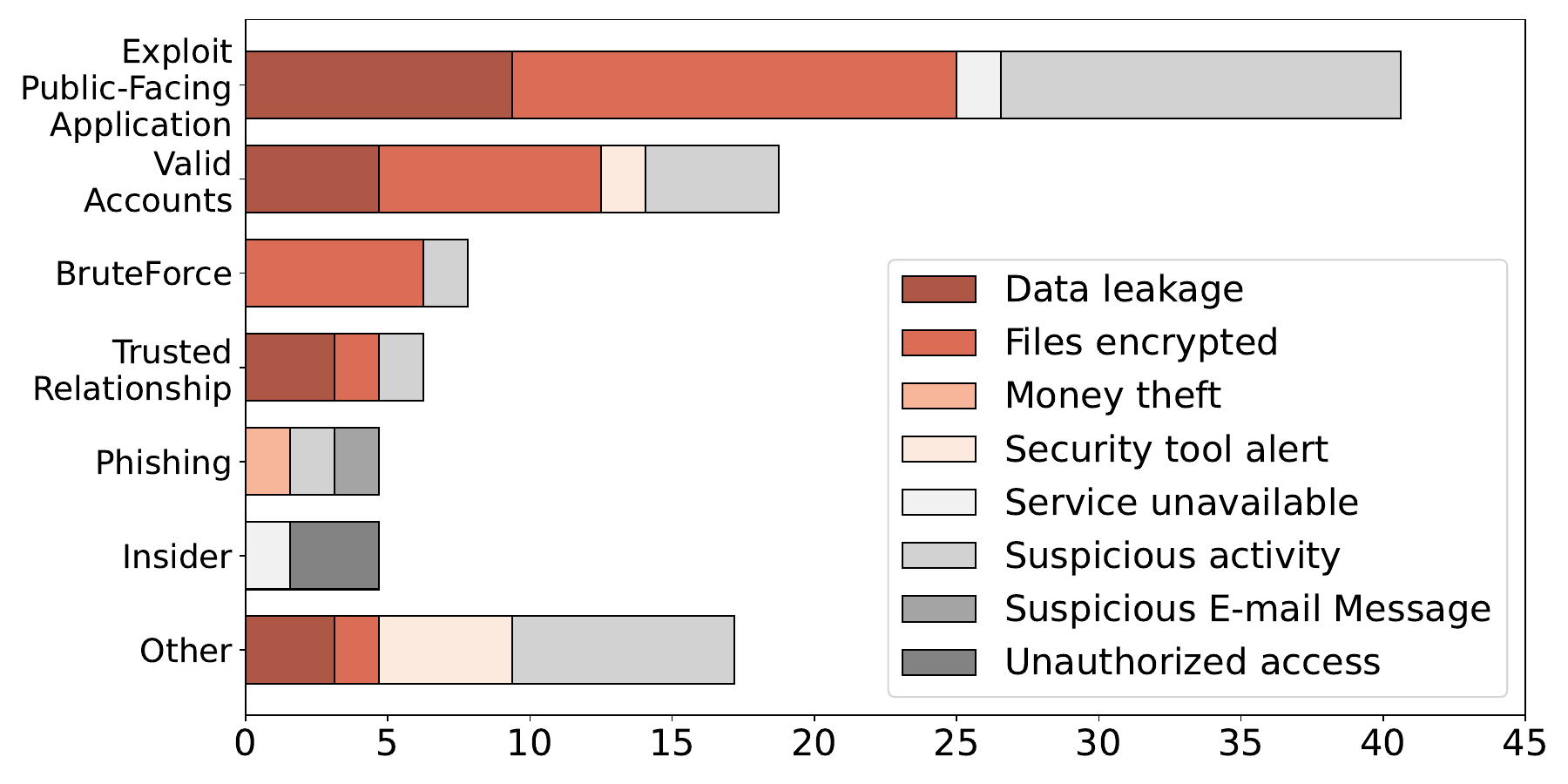}
  \vspace{-1em}
  \caption{Security Vulnerabilities Introduced by Various Attack Vectors.}
  \label{fig:Attack-Vectors}
  \vspace{-1em}
\end{figure}

\begin{figure}[!h]
  \centering
  \includegraphics[width=\linewidth]{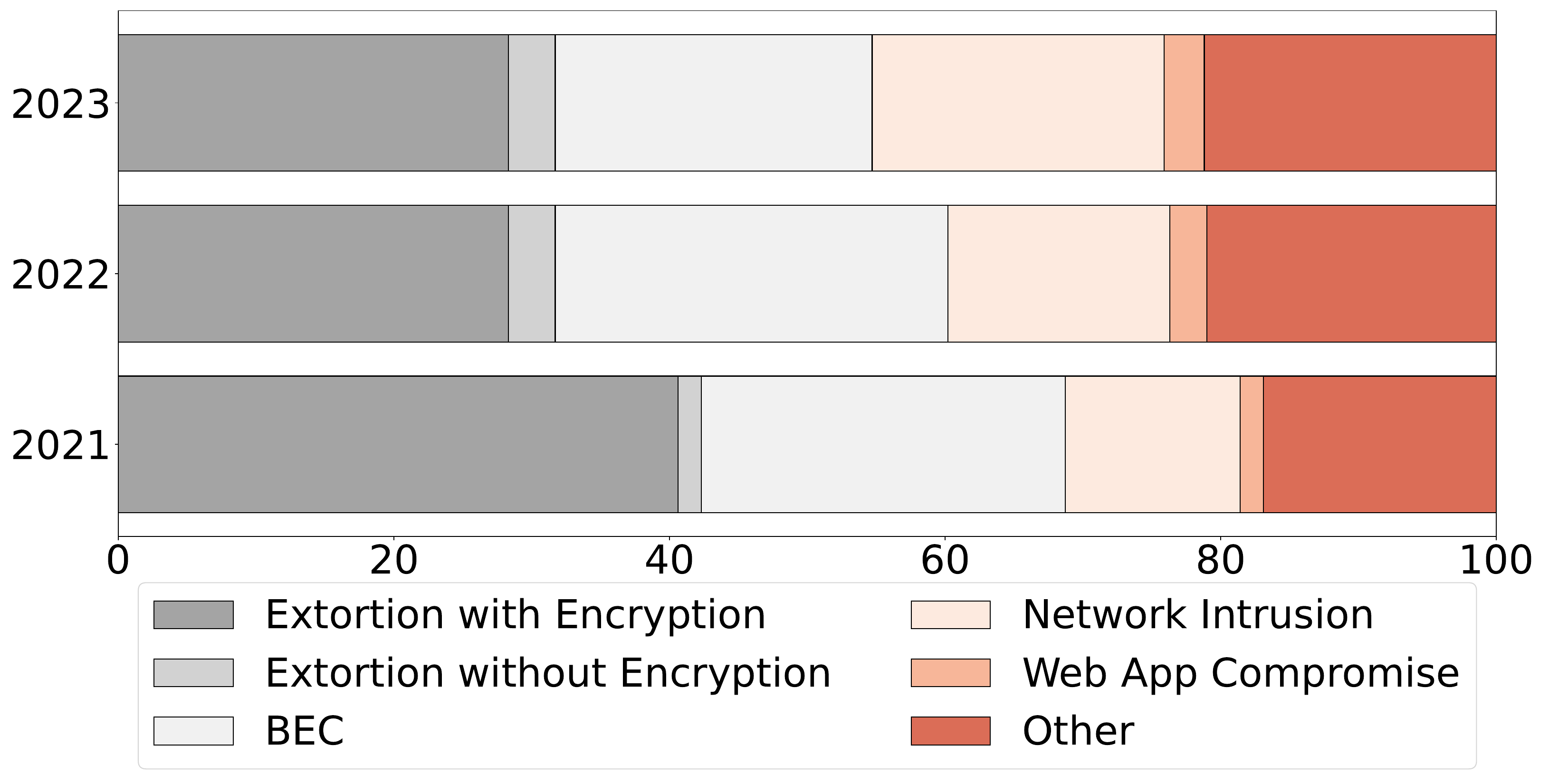}
  \vspace{-1em}
  \caption{Top 5 Investigation Types in Unit 42 IR Cases (2021-2023).}
  \label{fig:Investigation-Types}
  \vspace{-2em}
\end{figure}

%------------------------------------------------------------------------------------------------------------------
% \clearpage
\section{Case Discussion}\label{appendix:case}
\subsection{Case I (False IR Strategy)}\label{appendix:case1}
In Task 2 of the benchmark "ZGSF\_Linux 1", the "False Response Strategy" results in a typical failure. 
\begin{custombox}{\textbf{\textit{False IR Strategy (LLMs)}}}
Use the following methods to systematically identify potential flag information in the system:\\
1. Search for all possible flag files in the system:\\
\texttt{find / -type f -name "flag*" 2>/dev/null} \\
2. Inspect memory or processes for hidden information:\\
\texttt{strings /dev/mem | grep -i "flag"}\\
3. Analyze network traffic and scheduled tasks:\\
\texttt{tcpdump -A -i eth0 | grep "flag"}\\
4. Examine attack scripts and environment variables:\\
\texttt{env | grep "flag"} 
\end{custombox}

IR should adopt a \textbf{step-by-step clue analysis process} instead of relying solely on keyword searches, as some clues may lie beyond the files. This highlights a key limitation of LLMs in devising effective strategies for IR tasks. To address this, we introduced IRT in \tool{} to prioritize IR tasks. When the task lacks clear information, it will be executed sequentially in accordance with the order in the "Incident Response Procedures."
\begin{custombox}{\textbf{\textit{IRCopilot-Planner}}}
\textbf{\textit{IRT:}}\\
1. Incident Response Objectives (Linux) - [To-do]\\
\hspace*{1em}1.1 Hacker's IP address - (To-do)\\
\hspace*{1em}1.2 Flag 1 - (To-do)\\
\hspace*{1em}...\\
2. Incident Response Procedures - [To-do]\\
\hspace*{1em}2.1 Review Command History - (To-do)\\
\hspace*{1em}2.2 Investigate Sensitive Directories - (To-do)\\
\hspace*{1em}2.3 Analyze System Logs - (To-do)\\
\hspace*{1em}...\\
\rule{\textwidth}{0.4pt}\\
Task selection: \textbf{2.1 Review Command History}\\
1. View the command history of the current user.\\
\hspace*{1em}\texttt{\$ \textbf{history} \$}\\
2. Looking directly at the ".bash\_history" file may provide more information, especially after the user has logged out.\\
\hspace*{1em}\texttt{\$ \textbf{cat \textasciitilde/.bash\_history} \$}
\end{custombox}

%---------------------------------------------------------
\subsection{Case II (False Commands Generation)}\label{appendix:case2}
We present a specific example of a failure caused by the "False Command Generation" in Task 6 of "XuanJI\_Where-1S-tHe-Hacker" within the benchmark. When searching for hidden accounts, directly using the "net user" command fails to provide complete information. This case shows that LLMs may generate incorrect commands, which contradict facts and lead to task failures.
\begin{custombox}{\textbf{\textit{False Commands Generation (LLMs)}}}
Hidden Accounts Created by Hackers \\
Check the system account list: \\
1. Run \texttt{net user}  to view the list of accounts and check for any suspicious accounts.\\
2. Run \texttt{net user <username>} to view detailed information.
\end{custombox}

In \tool{}, the approach should involve first analyzing sub-tasks and then generating multiple solutions, specifically by \textbf{checking Regedit or using the WMIC command}.
\begin{custombox}{\textbf{\textit{IRCopilot-Generator}}}
To complete the task, we need to check in detail. Here are the specific steps and commands to perform this task:\\
1. Hidden accounts are typically recorded in the registry at the following location: \\
\hspace*{1em}\texttt{\$ \textbf{...\textbackslash SAM\textbackslash SAM\textbackslash Domains\textbackslash Account\textbackslash Users} \$} \\
This path stores all user SIDs and related information. \\
2. The WMIC can quickly list all accounts and their SIDs:\\
\hspace*{1em}\texttt{\$ \textbf{wmic useraccount get name,sid} \$}
\end{custombox}

%---------------------------------------------------------
\subsection{Case III (False Command Generation by \tool{})}\label{appendix:case3}
% We present an example of "XuanJI\_Where-1S-tHe-Hacker" in the benchmark, specifically Task 9, which involves identifying the moment the hacker reads the stored key from the file. This task remains unsolved by both LLMs and \tool{} due to hallucinations. The suggested Event IDs 4657 track general file access events, which are too broad and may not specifically capture cryptographic key operations. This can lead to missed or irrelevant logs, failing to pinpoint the exact moment of the key access. In contrast, \textbf{Event ID 5058 is the appropriate choice}, as it logs key file operations (\eg reading a stored cryptographic key) in the Windows Security event log, enabling accurate identification of the event.
We present an example from the "XuanJI\_Where-1S-tHe-Hacker" benchmark, specifically Task 9, which involves identifying the moment the hacker reads a stored key from a file. This task was ultimately marked as failed, and the failure is directly attributable to a critical sub-task where \tool{} failed to generate the correct forensic command. To diagnose the root cause of this error, we employed Gemini 2.5 pro, a model not used during the initial experiments, to evaluate the content generated by \tool{}.

\begin{custombox}{\textbf{\textit{IRCopilot-Planner}}}
1. Incident Response Objectives (windows) - [To-do]\\
\hspace*{1em}...\\
\hspace*{1em}1.9 Reserved key read time - (To-do)\\
\hspace*{2em}1.9.1 Investigate registry access logs - (To-do)\\
\hspace*{1em}...\\
\rule{\textwidth}{0.4pt}\\
Task selection: \textbf{1.9.1 Investigate registry access logs - (To-do)}\\
To execute this task, I will review the Windows registry access logs to identify any unauthorized access or modifications, focusing on key registry paths that store system configurations and credentials. This involves searching for unusual access patterns or changes that could indicate tampering or the use of a reserved key by the attacker.
\end{custombox}

\begin{custombox}{\textbf{\textit{IRCopilot-Generator}}}
...\\
1. Access Event Viewer:\\
\hspace*{1em}Open the Event Viewer to review security logs for registry access events. You can do this via the GUI or using PowerShell:\\
\hspace*{1em}\texttt{\$ \textbf{Get-EventLog -LogName Security | Where-Object \{ \$\_.EventID -eq 4657 \}} \$} \\
\hspace*{1em}Event ID 4657 is logged when registry value modified. Review these entries to identify any unauthorized changes.\\

2. Filter for Specific Registry Key Access:\\
\hspace*{1em}If you are looking for access to specific registry keys, filter the logs accordingly. For example, if you are looking for access to a specific registry key path:\\
\hspace*{1em}\texttt{\$ \textbf{Get-EventLog -LogName Security | Where-Object \{ \$\_.EventID -eq 4657 -and \$\_.Message -like "*RegistryPath*" \}} \$} \\
\hspace*{1em}Replace `"RegistryPath"` with the specific path.\\
...
\end{custombox}

% Our approach does not augment the existing knowledge base of LLMs. However, future research could develop a specialized knowledge base for LLMs tailored to IR to improve their performance in related tasks.
The evaluation revealed that the command suggested by \tool{} was fundamentally flawed. It proposed using Event ID 4657, which tracks general object access events. This ID is too broad and does not specifically capture cryptographic key operations, leading to missed or irrelevant logs that make it impossible to pinpoint the exact moment of key access. In contrast, the analysis by Gemini 2.5 pro confirmed that Event ID 5058 is the appropriate choice, as it specifically logs key file operations (e.g., reading a stored cryptographic key) in the Windows Security event log, which would have enabled accurate identification of the event.

\begin{custombox}{\textbf{\textit{Gemini 2.5 pro}}}
\#\# \textbf{Evaluation}\\
The model recommended the wrong Event ID (4657 instead of 5058). The suggested command checks for registry modifications, while the objective was to find the read time for a cryptographic key.\\

\#\# \textbf{Analysis of the Error}\\
1. \textbf{Over-generalization}: The model treated the task as a routine registry check, providing a generic but inaccurate answer.\\
2. \textbf{Ignoring Context}: It failed to recognize the keyword "Reserved key" and therefore missed its link to the specific cryptographic Event ID (5058).\\
3. \textbf{Logical Mismatch}: Recommending an ID for "modification" (4657) to accomplish a "read" task shows a lack of precision, even in a general context.
\end{custombox}

%---------------------------------------------------------
\subsection{Case IV (Key Information Ignored by \tool{})}\label{appendix:case4}
\begin{custombox}{\textbf{\textit{IRCopilot-Analyst}}}
...\\
\textit{Analysis}:\\
1. Attacker's IP: The traffic indicates communication from `192.168.20.1` to `192.168.20.144`.\\
2. Flag 1: The HTTP GET request to `\/flag1` resulted in a response containing `flag1\{Network@\_2020\_Hack\}`.\\
3. Webshell Connection: The POST request to `\/index.php? user-app-register` contains PHP code, suggesting a potential Webshell or malicious script.\\
4. Further Investigation: The POST request content involves PHP functions that manipulate server settings, potentially indicating a backdoor or malicious activity.\\

\textit{Updated IRT}:\\
1. Incident Response Objectives (linux) - [To-do]\\
\hspace*{1em}1.1 Attacker's IP - (192.168.20.1)\\
\hspace*{1em}1.2 Web system administrator plain text password - (To-do)\\
\hspace*{1em}1.3 URL of the first Webshell connection - (/index.php?user-app-register)\\
\hspace*{1em}1.4 Webshell connection parameter - (To-do)\\
\hspace*{1em}1.5 Subsequent upload of Webshell name - (To-do)\\
\hspace*{1em}1.6 Flag 1 - (flag1\{Network@\_2020\_Hack\})\\
...
\end{custombox}

We present an example from the "ZGSF\_Linux2" benchmark, specifically Task 4, which involves identifying the webshell connection parameter. This task was ultimately marked as failed, and the failure is directly attributable to Key Information Ignored, as the model overlooked the relevant data in the request body.

This failure stemmed from the model's inability to parse the request's basic structure. It focused exclusively on the complex PHP payload (the "value"), thereby ignoring the parameter name Network2020 (the "key"), which was the actual answer. A successful analysis would have first identified the request's key=value format to extract the parameter before proceeding to analyze the payload's content.

\begin{custombox}{\textbf{\textit{Gemini 2.5 pro}}}
\#\# \textbf{Evaluation}\\
The model failed to extract the webshell connection parameter (Network2020) from the POST request. It incorrectly marked this item as (To-do), while the correct answer was explicitly available in the key=value structure of the request body.\\

\#\# \textbf{Analysis of the Error}\\
1. \textbf{Over-focus on Payload}: The model dedicated all its analytical resources to the complex PHP malware (the value), completely ignoring the parameter name (the key) that was used to carry it.\\
2. \textbf{Ignoring Structural Context}: It failed to parse the fundamental key=value format of the POST request body, thereby missing the entry point parameter for the entire attack.\\
3. \textbf{Pattern Matching Failure}: The parameter name Network2020 does not fit the typical pattern of webshell parameters like cmd or eval, which likely caused the model's heuristics to overlook it.
\end{custombox}

%---------------------------------------------------------
\subsection{Case V (Limitation in Reverse)}\label{appendix:case5}
Here, we offer an example task highlighting the limitations of LLMs and \tool{}: the Squid Game malicious file analysis challenge from TryHackMe. This challenge demands sophisticated dynamic and static analysis, encompassing data flow analysis, VBA macro handling, obfuscated content decoding, and the seamless integration of multiple tools.

\begin{custombox}{\textbf{\textit{Results of Oledump}}}
ubuntu@ip-10-10-114-110:\textasciitilde/Desktop/maldocs\$ oledump.py attacker1.doc\\
 1: \hspace*{2em} 114 '\textbackslash x01CompObj'\\
...\\
 8: M\hspace*{1em}9852 'Macros/VBA/ThisDocument'\\
...
\end{custombox}

% ubuntu@ip-10-10-114-110:~/Desktop/maldocs$ oledump.py attackerl.doc
%  1:       114 '\x01CompObj'
%  2:      4096 '\x05DocumentSummaryInformation'
%  3:      4096 '\x05SummaryInformation'
%  4:     13859 '1Table'
%  5:     33430 'Data'
%  6:       365 'Macros/PROJECT'
%  7:        41 'Macros/PROJECTwm'
%  8: M    9852 'Macros/VBA/ThisDocument'
%  9:      5460 'MacroS/VBA/_VBA_PROJECT'
% 10:       513 'Macros/VBA/dir'
% 11:       306 'MsoDataStore/CYOXGNI0AUKW0IIS2BKIDD==/Item'
% 12:       341 'MsoDataStore/CYOXGNI0AUKW0IIS2BKIDD==/Properties'
% 13:      4096 'WordDocument'

\begin{custombox}{\textbf{\textit{LLMs}}}
oledump.py -s 8 -v attacker1.doc\\
oledump.py -s 1 -d attacker1.doc
\end{custombox}

The output from these two commands generates voluminous, obfuscated data streams that are challenging for current LLMs to parse, often necessitating intricate manual analysis. We anticipate this will be a key challenge addressed by the next generation of LLM-based software analysis.
%---------------------------------------------------------
\newcolumntype{L}[1]{>{\raggedright\arraybackslash}p{#1}}
\begin{table*}[]
	\centering
	\caption{Difficulty Levels and Descriptions for Incident Response Tasks}
	\fontsize{6}{8}\selectfont
	\resizebox{\textwidth}{!}{
		\begin{tabular}{c|l|L{5cm}|L{5cm}}
			\hline
			\textbf{Difficulty Score} & \textbf{Difficulty Level} & \textbf{Description} & \textbf{Example} \\
			\hline
			\multirow{2}{*}{1} & \multirow{2}{*}{Easy} & Steps are clear and simple, can be completed quickly. Typically no extra tools are required; relies on common system utilities. & Run \texttt{netstat} or \texttt{tasklist} on a Windows host to list active network connections and processes. \\
			\hline
			\multirow{2}{*}{2} & \multirow{2}{*}{Easy} & Requires basic security knowledge and a brief period of analysis; some scenarios may need specialized tools. & Use a PowerShell script or \texttt{grep} to filter SSH login failures from system logs over the past hour and export the list of suspicious IPs. \\
			\hline
			\multirow{2}{*}{3} & \multirow{2}{*}{Easy} & Combines elements of levels 1 and 2 or requires some IR theory; involves static analysis of malicious code and basic programming skills. & Extract a \texttt{Base64}-encoded suspicious script and use a static analysis tool to inspect its unpacked core functionality. \\
			\hline
			\multirow{2}{*}{4} & \multirow{2}{*}{Medium} & Builds on the above, requiring writing scripts or complex commands, and the ability to learn and adapt quickly. & Write a script to extract key events from multiple log sources (such as operating system logs, security audit logs, and network device logs) and analyze them. \\
			\hline
			\multirow{2}{*}{5} & \multirow{2}{*}{Medium} & Targets are diverse and hard for a single analyst to cover; requires team collaboration and in-depth study of security principles and code vulnerabilities. & Cross-correlate anomalous login events with large-scale outbound traffic events to pinpoint potential lateral movement paths. \\

			\hline
			\multirow{2}{*}{6} & \multirow{2}{*}{Medium} & Demands systematic learning and long-term effort on specific targets, rigorously testing an analyst’s deep knowledge and hands-on experience. & In a Windows environment, batch-extract and parse complex event IDs (e.g., 4624, 4688, 5156) from security logs, standardize key fields, and correlate across hosts using user, process, and network data. \\

			\hline
			\multirow{2}{*}{7} & \multirow{2}{*}{Hard} & Requires continuous learning and practice to keep pace with evolving techniques and progressively meet higher difficulty challenges. & In-depth static and dynamic analysis. \\

			\hline
			\multirow{2}{*}{8} & \multirow{2}{*}{Hard} & Attack has penetrated the application-framework layer; demands comprehensive analysis and hardening of the entire framework. & Detect and remove memory-resident backdoors (memory horses) and patch affected framework components to secure inter-component calls. \\

			\hline
			\multirow{2}{*}{9} & \multirow{2}{*}{Hard} & Attack has reached the operating-system level; requires integrating multi-source data for systematic analysis, possibly involving 0day research/exploit. & On a compromised Windows Server, detect and track implanted cryptomining or ransomware processes, correlate event logs, network traffic, and registry changes, then locate and remove persistence mechanisms. \\

			\hline
			\multirow{2}{*}{10} & \multirow{2}{*}{Hard} & Targets low-level architecture (e.g., CPU kernel or chip); involves reverse engineering, large-scale binary analysis, and hardware-level security research. & Reverse-engineer firmware or kernel modules to identify microarchitecture or hardware side-channel vulnerabilities, then design microcode or firmware-level mitigations or patches. \\

			\bottomrule
		\end{tabular}
	}
	\label{tab:difficulty_levels}
\end{table*}

%---------------------------------------------------------
\begin{table*}[]
	\centering
	\caption{Summarized 27 types of sub-tasks in the proposed incident response benchmark.}
	\fontsize{14}{16}\selectfont
	\resizebox{\textwidth}{!}{
		\begin{tabular}{l|l|l}

			\hline
			\textbf{Phase}                       & \textbf{Technique}                      & \textbf{Description}                                                                                                  \\ \hline
			\multirow{7}{*}{\textbf{Detection}}  & System Information Gathering            & Includes operating system identification, network configuration analysis, hardware information gathering, etc.        \\
			                                    & Open Port Identification                & Detect open network ports on the target system.                                                                       \\
				                                & Service Enumeration                     & Identify and analyze running services to uncover version details and vulnerabilities.                                 \\
			                                    & Directory Inspection                    & Examine key directories and hidden files for unusual activity.                                                        \\
			                                    & Account Security Review                 & Audit user account permissions and identify unauthorized accounts or backdoors.                                       \\
			                                    & File Integrity Check                    & Monitor file hashes to detect unauthorized changes.                                                                   \\
			                                    & Other Detections                        & Inspect other vulnerable areas of the system, such as the registry, etc.                                              \\ \hline             
			\multirow{15}{*}{\textbf{Response}}  & Historical Command and Behavior Analysis& Review user commands and system behaviors to detect abnormal operations.                                              \\
			                                    & Permission Review and Management        & Audit system and application permissions to enforce least-privilege principles and manage risky permissions.          \\
			                                    & File Analysis                           & Inspect system files and source code for vulnerabilities or malicious content.                                        \\
			                                    & Malicious File Handling                 & Identify, isolate, and remove malicious files to prevent further damage.                                              \\
			                                    & Startup Item Analysis                   & Review startup items for unauthorized programs or scripts.                                                            \\
			                                    & Scheduled Task Analysis                 & Analyze the system's scheduled tasks settings to identify possible malicious or planned tasks.                        \\
			                                    & Anomaly Behavior Response               & Respond to abnormal system or user behaviors to contain potential threats.                                            \\
			                                    & Memory and Process Analysis             & Examine memory and processes to identify abnormal or malicious activity.                                              \\
			                                    & Malicious Process Handling              & Terminate malicious processes to mitigate ongoing threats.                                                            \\
			                                    & System Log Analysis                     & Analyze system logs for signs of compromise, unauthorized access, or suspicious activities.                           \\
			                                    & Application Log Analysis                & Review application logs for exploitation attempts or unusual behavior.                                                \\
			                                    & Network Traffic Analysis                & Analyze network traffic to identify suspicious communication or data exfiltration.                                    \\
			                                    & Risky IP Management                     & Block or monitor traffic from known malicious or suspicious IP addresses.                                             \\
			                                    & Database Analysis                       & Analyze databases for security vulnerabilities, data leaks, or unauthorized access.                                   \\
			                                    & Other Responses                         & Conduct additional response activities, such as analyzing virtualization environments or reviewing container security.\\ \hline
			\multirow{5}{*}{\textbf{Recovery}}   & System Recovery                         & Restore the system to a stable state after failures, malware, or misconfigurations.                                   \\
			                                    & Data Recovery                           & Recover lost or corrupted data from backups or damaged media.                                                         \\
			                                    & Service Recovery                        & Restore key services and applications to minimize downtime.                                                           \\
			                                    & Vulnerability Patching                  & Apply patches to fix vulnerabilities and prevent recurrence of attacks.                                               \\
			                                    & Other Recoveries                        & Additional recovery methods, such as network recovery and permission reset, to address various aspects of system restoration.
			\\ \bottomrule
		\end{tabular}
	}
	\label{appendix:benchmark}
\end{table*}